\documentclass[reprint,aps,prb,twocolumn,nofootinbib,floatfix,showpacs]{revtex4-1}
\usepackage{graphicx}
\usepackage{caption}
\usepackage{color}
\usepackage{amsmath}
\usepackage{amssymb}
\usepackage{comment}
\usepackage{hyperref}
\usepackage{float}
\usepackage{setspace}
\usepackage{subfig}
\usepackage{tikz}
\usetikzlibrary{shapes,arrows}
\usepackage{titlesec}

\def\Im{{\mbox{Im}}}

\definecolor{scarred}{rgb}{0.75,0.0,0.0}

\makeatletter
\renewcommand\@makecaption[2]{%
  \par
  \vskip\abovecaptionskip
  \begingroup
   \small\rmfamily
    \begingroup
     \samepage
     \flushing
     \let\footnote\@footnotemark@gobble
     \@make@capt@title{#1}{#2}\par
    \endgroup
  \endgroup
  \vskip\belowcaptionskip
}
\makeatother

\begin{document}
\title{Temperature and doping induced instabilities of the repulsive Hubbard model on Lieb lattice}
\author{Pramod Kumar}
\author{Tuomas I. Vanhala}
\author{P\"aivi T\"orm\"a}
\affiliation{COMP Centre of Excellence, Department of Applied Physics, Aalto University,
Helsinki, Finland}
\begin{abstract}
The properties of a phase at finite interactions can be significantly influenced by the underlying dispersion of the non-interacting Hamiltonian. We demonstrate this by studying the repulsive Hubbard model on the $2$D Lieb lattice, which has a flat band for vanishing interaction $U$. We perform real-space dynamical mean-field theory calculations at different temperatures and dopings using a continuous time quantum Monte Carlo impurity solver. Studying the frequency dependence of the self-energy, we find that a non-magnetic metallic region at finite temperature displays non-Fermi liquid behavior, which is a concomitant of the flat band singularity. At half-filling, we also find a magnetically ordered region, where the order parameter varies linearly with the interaction strength, and a strongly correlated Mott insulating phase. The double occupancy decreases sharply for small $U$, highlighting the flat band contribution. Away from half-filling, we observe the stripe order, i.e. an inhomogeneous spin and charge density wave of finite wavelength which turns into a sub-lattice ordering at higher temperatures. 
\end{abstract}
\pacs{ Strongly correlated electron systems, Non-Fermi-liquid ground state, Cold atoms}
\maketitle
\section{Introduction}
\label{sec:intro}
Singularities in the non-interacting density of states (DOS), such as a Van Hove singularity or a flat band, inflate the instabilities towards various ordered states at finite interactions. Such singularities can affect magnetically ordered states \cite{PhysRevLett.78.1343} and enhance superconductivity ~\cite{PhysRevLett.56.2732}, and have substantial consequences in two dimensions ($2$D)~\cite{PhysRevB.80.245112}. A flat band, which can be represented by $\delta$-function in the energy spectrum, is even more singular than a Van Hove singularity and leads to correlation induced novel phases which are qualitatively different from the phases appearing in presence of a Van Hove singularity~\cite{doi:10.1143/PTP.99.489}. Influence of the flat bands or quasi-flat bands on different emergent novel phases of interacting lattice fermions, such as ferromagnetism~\cite{PhysRevLett.62.1201, PhysRevA.80.063622, PhysRevB.93.235143, PhysRevLett.88.127202}, flat band superfluidity~\cite{PhysRevLett.117.045303,Peotta2015,PhysRevB.95.024515}, high T$_c$ superconductivity of electron-doped compounds~\cite{Khodel2015}, non-Fermi-liquid behavior~\cite{PhysRevLett.115.156401, hausoel2017local} and topological phases~\cite{PhysRevA.83.063601} have been explored theoretically. Experimentally, effects of flat bands have been reported in various real materials such as tetragonal  La$_4$Ba$_2$Cu$_2$O$_{10}$\cite{doi:10.1143/PTP.99.489}, LaCo$_5$ and CePt$_5$~\cite{PhysRevB.91.165137}, and can be realized using ultra cold atoms~\cite{Taiee1500854,Greif953,RevModPhys.80.885,doi:10.1146/annurev-conmatphys-070909-104059}, where lattice geometry, and thus the singularities, can be well controlled. Breakdown of the Fermi-liquid (FL) theory in a class of metallic systems~\cite{0034-4885-68-10-R02,PhysRevB.88.195120}, as seen in transport properties, can be attributed to such singularities in the DOS~\cite{PhysRevB.80.245112}. The diverging DOS can significantly affect the stripe order, which appears at finite doping, and has been extensively studied in the context of the pseudo-gap region of cuprate high temperature superconductors~\cite{VOJTA2012178}.

Flat bands can be realized using different lattice model Hamiltonians \cite{PhysRevLett.62.1201, doi:10.1143/PTP.99.489, PhysRevLett.106.236803, PhysRevB.88.235303}. A simple model displaying a flat band is the Lieb lattice, a bipartite lattice, as shown in Fig.~\ref{fig1}. The non-interacting $2$D Lieb lattice has been realized using ultra cold atoms~\cite{Taiee1500854,PhysRevLett.118.175301}, photonic lattices~\cite{PhysRevLett.114.245503,PhysRevLett.114.245504} and also electronically~\cite{Slot2017,Drost2017}. To explore flat band ferromagnetism, the repulsive Hubbard model on the $2$D Lieb lattice has previously been studied using real space dynamical mean theory (R-DMFT) combined with a numerical renormalization group (NRG) impurity solver at half-filling and zero temperature~\cite{PhysRevA.80.063622}. The findings are in agreement with the Lieb theorem \cite{PhysRevLett.62.1201}, which states that the ground state of the repulsive Hubbard model on a bipartite lattice in any dimension with an unequal number of sites in each sub-lattice must have a non-zero net magnetic moment at half-filling. Finite temperature effects on an anisotropic three dimensional Lieb lattice have been studied using R-DMFT+NRG~\cite{PhysRevA.91.063610}. For a weak interplane coupling, the authors find remnants of $2$D Lieb lattice behavior in different physical observables. For a specific choice of parameters, the double occupancy increases with increasing temperature violating the FL theory. Finite size determinantal quantum Monte-Carlo has also been employed to explore the flat band contribution to the interaction-induced magnetic ordering~\cite{PhysRevB.93.235143} at half-filling. The magnetic behavior is characterized by the local moment and the real space spin correlations. The approach, however, suffers from a sign problem away from half-filling. \\
\indent In general, as stated by the Mermin-Wagner-Hohenberg theorem~\cite{PhysRevLett.17.1133,PhysRev.158.383}, a continuous symmetry cannot be spontaneously broken at finite temperature in $2$D systems. However, one can define a finite temperature scale related to the development of short range magnetic order~\cite{PhysRevLett.104.066406}. Such a scale has well defined signatures in physical observables and has been observed in fermionic cold atom experiments recently~\cite{Cheuk1260,Parsons1253,mazurenko2017cold,Greif1236362,PhysRevLett.118.170401,Brown1385}. As DMFT neglects long range fluctuations, it breaks the Mermin-Wagner-Hohenberg theorem and formally allows a long range magnetization to develop at a temperature $T_N^d$~\cite{PhysRevB.95.235109}. This temperature scale, while it does not correspond to a true phase transition point, gives an estimate of the temperature where short range magnetic correlations become significant. In the present paper, our main goal is to elucidate the influence of a flat band on the breakdown of FL theory in the non-magnetic metallic region appearing at a finite temperature, finite interactions and half-filling. We observe various intriguing regimes, such as a magnetically ordered state where local magnetization scales linearly with the interaction strength, and a non-magnetic Mott insulator. We also study the stripe order, evident away from half-filling, which is naturally viable with R-DMFT. R-DMFT, an extension of DMFT, has been successfully employed to study e.g. topological systems, interfaces and trapped correlated systems \cite{PhysRevB.94.115161,PhysRevB.87.224513,PhysRevB.83.054419}. Here we apply R-DMFT coupled with a continuous time interaction expansion (CTINT) quantum Monte Carlo \cite{RevModPhys.83.349} impurity solver.\\
\indent The paper is structured as follows. We first introduce the dimerized Hubbard model on the Lieb lattice, followed by the formalism of the real space dynamical mean field theory,
which is used to incorporate the effects of correlations and quantum fluctuations. In section III A, we present the phase diagram in the presence of interactions, which is the central result of this work. We show the flat-band-induced non-Fermi liquid behavior in section III B. In sections III C and III D, we discuss the behavior of the double occupancy and the dimerization effects on the physical observables, respectively. In section III E, we present our findings about the doping effects and discuss the observed stripe ordering and also show a phase diagram for the doped case at a fixed finite temperature.

\section{The model and the formalism}
\subsection{The dimerized Hubbard model on the Lieb Lattice}
\label{sec:model}
\begin{figure}

\includegraphics[scale=0.7,height=0.18\textheight,width=0.45\textwidth]{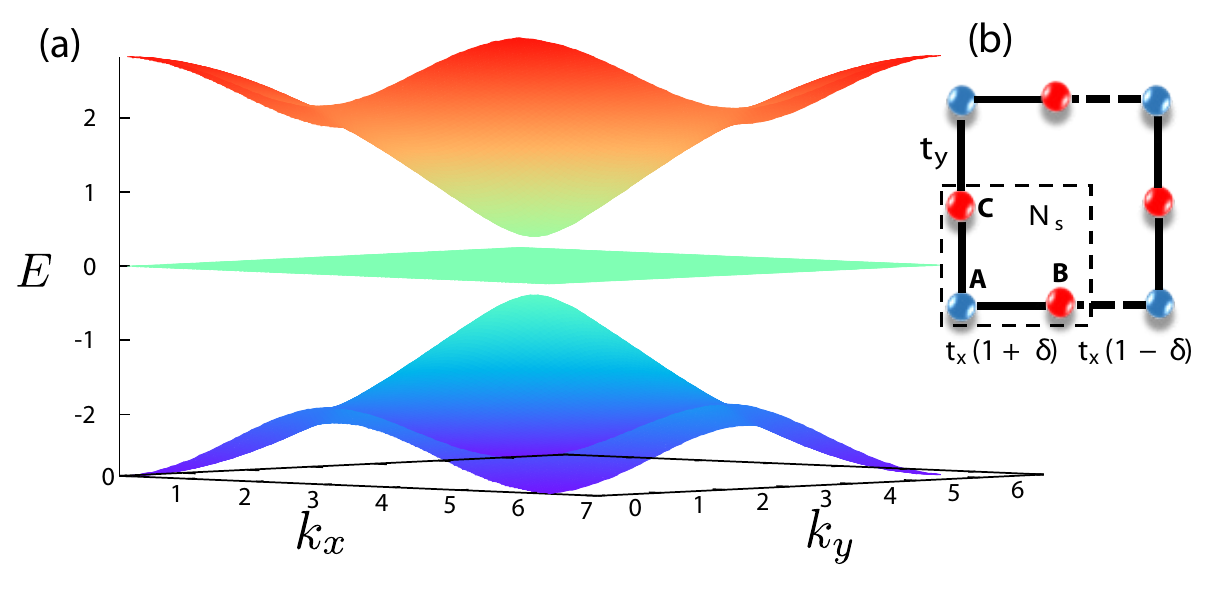}
\includegraphics[scale=0.7,height=0.25\textheight,width=0.4\textwidth]{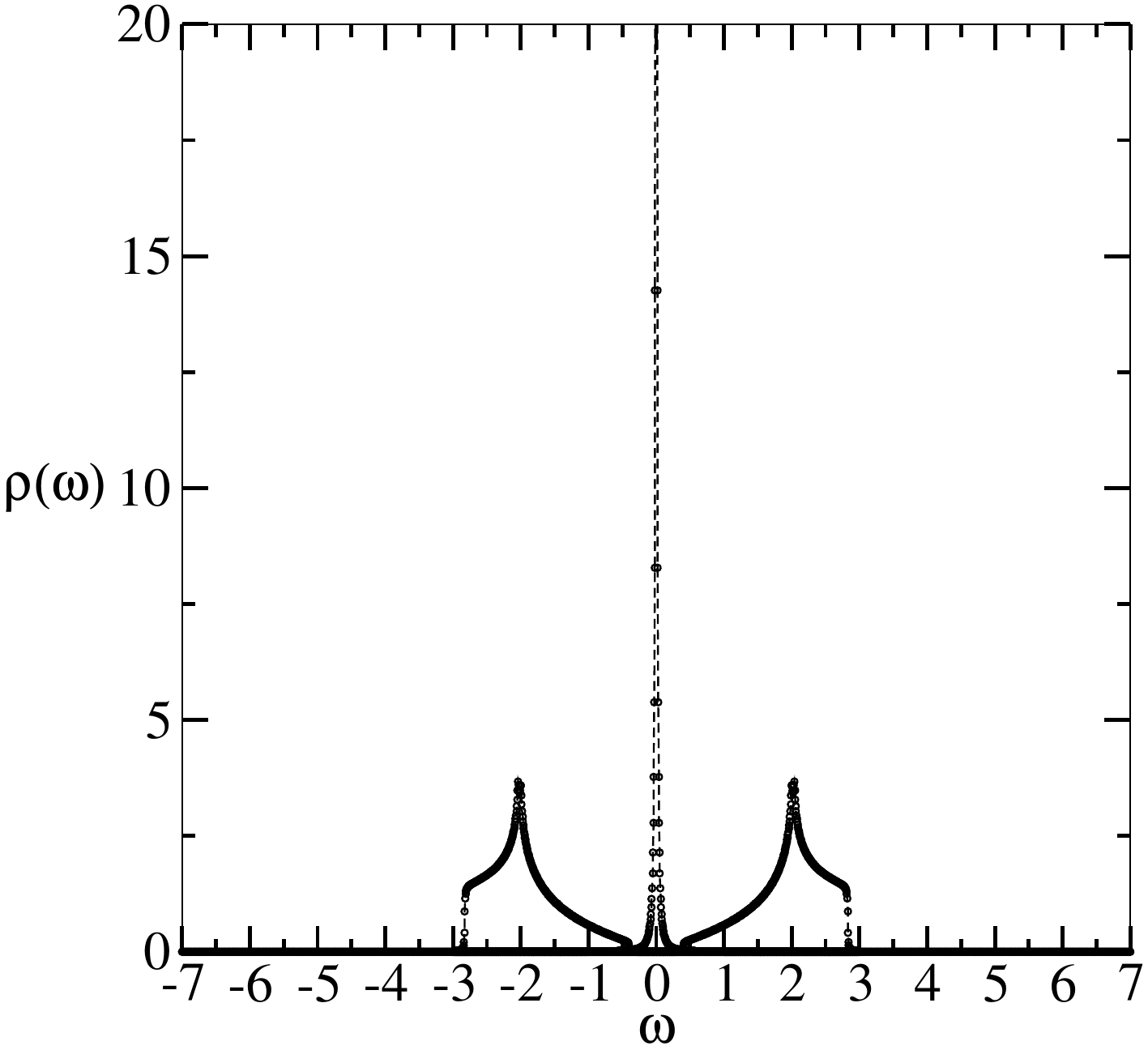}
\caption{Upper panel : (a) The non-interacting dispersion of the Lieb lattice for a finite dimerization $\delta>0$. There are upper and lower dispersive bands and an isolated dispersionless band with zero energy, i.e. $E(k_x,k_y)=0$, for all lattice momenta $k_x$ and $k_y$. (b) A schematic representation of the Lieb lattice, where blue dots mark the $A$ type of sites while red dots mark the $B$ and $C$ types. The rectangle drawn with a dashed line is the smallest possible unit cell that captures the magnetic ordering emerging for a finite interaction strength $U$. Lower panel:  Non-interacting density of states corresponding to the dispersion shown above.}
\label{fig1}
\end{figure}
The Lieb lattice in two dimensions is characterized by a three-site unit cell where the sites are labeled as $A$, $B$ and $C$, as shown in Fig.~\ref{fig1}. The Hamiltonian of the Hubbard model on this lattice can be expressed as $H=H_{\text{t}}-\mu N+H_U$, where the first term is the tight-binding part represented in standard second quantized notation as
\begin{align}
H_{\text{t}}=&-\sum_{j,\sigma}\Big[(t_x (1+\delta) c_{A,j,\sigma}^\dagger c^{\phantom{\dagger}}_{B,j,\sigma}+t_y c_{A,j,\sigma}^\dagger c^{\phantom{\dagger}}_{C,j,\sigma} 
+ \text{h.c.}) \nonumber \\ 
&+(t_x (1-\delta) c_{A,j,\sigma}^\dagger c^{\phantom{\dagger}}_{B,j-\hat{x},\sigma}+t_y c_{A,j,\sigma}^\dagger c^{\phantom{\dagger}}_{C,j-\hat{y},\sigma} 
+ \text{h.c.}) \Big] \label{eq1},
\end{align}
where $c_{A(B/C),j,\sigma}^\dagger$ is the creation operator corresponding to the site $A(B/C)$ for the unit cell at $j = (x, y)$.  $\hat{x} = (1, 0)$ and $\hat{y} = (0, 1)$ are the unit vectors. The first line corresponds to the intra-unit-cell hoppings and the rest represents hopping between neighboring unit cells. In this work we set $t_x=t_y=t$, and tune the $x$-directional hoppings via the dimerization parameter $\delta$. Such dimerization leads to the isolated flat band as shown in Fig.~\ref{fig1}(a) and can be used to tune the weight of the flat band in real-space. The second term $\mu N$ of the full Hamiltonian is the chemical potential, where the total particle number is $N=\sum_{s, j,\sigma } c_{s,j,\sigma}^\dagger c^{\phantom{\dagger}}_{s,j,\sigma}$, and $s=A,B$ or $C$. The last term is the on-site Hubbard interaction which can be defined as
\begin{equation}
H_U=U\sum_{s,j} (n_{s,j,\uparrow}-\frac{1}{2})(n_{s,j,\downarrow}-\frac{1}{2}),
\label{eq_Hu}
\end{equation}
where $U>0$ is the interaction strength. 

The eigenvalues of the non-interacting Hamiltonian $H_t$ can be given as
\begin{align}
&E_{\pm}=\pm \sqrt{|\Delta_x|^2+|\Delta_y|^2} \nonumber \\ 
&E_{0}=0 \label{eq2},
\end{align}
where $\Delta_x=2t_x\cos^2\frac{k_x}{2}+i \ 2t_x \delta \sin^2\frac{k_x}{2}$, $\Delta_y=2t_y\cos^2\frac{k_y}{2}$, $k_x=2\pi p/N (p=1,..,N)$ and $k_y=2\pi q/M (q=1,..,M)$, giving rise to a three band structure (Fig.~\ref{fig1}). $E_{\pm}$ are the eigenvalues for the upper and lower bands, respectively, and $E_0$ corresponds to the flat (non-dispersive) band of the Lieb lattice~\cite{PhysRevA.83.063601}. $E_{\pm}$ acquires the semi-metal dispersion for $\delta=0$ touching the flat band at the point $(k_x,k_y)=(\pi,\pi)$. The corresponding eigenfunctions for different bands can be given as
\begin{align}
&\psi_{\pm}=\frac{1}{2}(\pm c_{A,k,\sigma}^\dagger+\frac{\Delta_x}{\sqrt{|\Delta_x|^2+|\Delta_x|^2}}c_{B,k,\sigma}^\dagger \nonumber \\
&+\frac{\Delta_y}{\sqrt{|\Delta_x|^2+|\Delta_y|^2}}c_{C,k,\sigma}^\dagger)|0\rangle \nonumber \\
&\psi_0=\frac{1}{\sqrt{|\Delta_x|^2+|\Delta_y|^2}}(\Delta_y^* c_{B,k,\sigma}^\dagger-\Delta_x^*c_{C,k,\sigma}^\dagger )|0\rangle \label{eq3},
\end{align}
where $c_{A(B/C),k,\sigma}^\dagger=\frac{1}{\sqrt{MN}}c_{A(B/C),j,\sigma}^\dagger e^{ik \cdot j}$. With the tuning of the dimerization parameter $\delta$, the weight of the flat-band can be tuned between $B$ and $C$ sites and thus the flat-band contribution can be distributed between the local quantities, as discussed in section \ref{sec:dimer}. In the next section, we will discuss our implementation of the dynamical mean-field theory for this model.

\subsection{Real-space dynamical mean field theory}
\label{sec:dmft}

To investigate the effects of correlations on the Lieb lattice, we have employed real-space dynamical mean-field theory (R-DMFT) which captures the simple magnetic states as well as the stripe ordered states with wavelengths more than two sites appearing in the doped regime \cite{PhysRevB.89.155134,vanhala2017dynamical}. DMFT maps a lattice problem to an effective single impurity problem taking into account the lattice effects in a self-consistent manner~\cite{RevModPhys.68.13}.
A central quantity is the self-energy $\Sigma_{ij\sigma}(i \omega_n)$, where $i$ and $j$ index the lattice sites, $\sigma$ is a spin index and $\omega_n=\pi(2n+1)T$, where $T$ is the temperature, are the Matsubara frequencies. Within single-site DMFT the self-energy is assumed to be local to each site $i$ and uniform over the whole lattice, so that $\Sigma_{ij}(i \omega_n) \sim \delta_{ij}\Sigma(i \omega_n)$. For magnetized states, however, the uniformity assumption breaks, as the magnetization can be different for different lattice sites. To study such states we thus use R-DMFT where the self-energy is still local but varies spatially, i.e. $\Sigma_{ij,\sigma}(i\omega_n)=\Sigma_{\sigma}^{i}(i\omega_n)\delta_{ij}$~\cite{1367-2630-10-9-093008}. 

In practice the self-energy is allowed to vary spatially within an enlarged unit cell, which can be larger than the basic three-site unit cell of the Lieb lattice. At half-filling, it is expected that the three-site unit cell (Fig. \ref{fig1}) is sufficient to investigate the interaction-induced order parameters, while larger magnetic unit cells should be considered to capture the stripe order appearing in the doped case. In the doped regime we have considered unit cells with numbers of sites up to $36$, where the three-site cell is stacked $12$ times linearly.

More rigorously, the R-DMFT method for a given unit cell can be described as follows. The local Green's function of the lattice system limited to a single unit cell can be calculated as
\begin{equation}
  \mathbf{G}_\sigma(i\omega_n)= \frac{1}{N_{k}}\sum_{\mathbf{k}} \left( \mathbf{G}_{\mathbf{k} \sigma}^0(i\omega_n)^{-1}-\mathbf{\Sigma}_\sigma(i\omega_n) \right)^{-1}, \label{eq4}
\end{equation}
where the bold quantities are matrices whose dimension equals the number of sites within the unit cell and $N_k$ is the number of $k$- points. Thus the matrix element $\mathbf{G}_{\sigma}(i \omega_n)_{ij}$ is the Green's function between sites $i$ and $j$ of the unit cell. The non-interacting Green's function $\mathbf{G}_{\mathbf{k} \sigma}^0(i\omega_n)^{-1}=(\mu_\sigma+i\omega_n)\mathbf{1}-\mathbf{T}_{\mathbf{k}}$, where $\mathbf{1}$ is the unit matrix and $\mathbf{T}_{\mathbf{k}}$ is the superlattice Fourier transform of the hopping matrix. This equation has exactly the same form as the coarse graining relation of the cellular DMFT \cite{RevModPhys.77.1027}. However, in the R-DMFT case the self-energy is assumed to be diagonal in the site indices, even though it can be different for different sites.

For each site $i$ in the (enlarged) unit cell, there is an effective single impurity Anderson model, which is defined by the dynamical Weiss mean-field
\begin{equation}
\mathcal{G}_{\sigma}^{i}(i\omega_n)^{-1}=(\mathbf{G}_{\sigma}(i\omega_n)_{ii})^{-1}+\Sigma^{i}_{\sigma}(i\omega_n)_{ii}.\label{eq5}
\end{equation}
Given the Weiss function $\mathcal{G}_{\sigma}^{i}$ for all $i$, we calculate the self-energy of each of the impurity problems using a continuous time quantum Monte-Carlo (CTINT) algorithm~\cite{RevModPhys.83.349}. These new self-energies are then used again in equation \ref{eq4} and the process is iterated until a converged solution is found.

In the half-filled case, we define the local magnetization for different sites in the unit cell as
\begin{equation}
m_{A(B/C)}=n_{A(B/C),\uparrow}-n_{A(B/C),\downarrow}, \label{eq6}
\end{equation}
where $n_{A(B/C),\sigma}=G_{A(B/C),\sigma}(\tau \rightarrow 0^-)$ is the density of spin-$\sigma$ particles at the chosen site. The striped magnetic order in the doped regime can be observed using
\begin{equation}
 m(r_x,r_y)=|n_{r_x,r_y,\uparrow}-n_{r_x,r_y,\downarrow}|\label{eq7}
 \end{equation}
where $r_{x(y)}$ are the positions of the sites in the unit cell. Similarly, we denote the total density as
\begin{equation}
 n(r_x,r_y)=  n_{r_x,r_y,\uparrow}+n_{r_x,r_y,\downarrow}\label{eq8}
 \end{equation}
For the half-filled case, we also calculate the double occupancy $D=\langle n_{\uparrow}n_\downarrow \rangle$ to study the correlation effects in the presence of the flat band.
\section{Results and discussion}
\label{sec:result}
The main purpose of this work is to explore the influence of the flat band, present in the dispersion of the non-interacting Lieb lattice, on the different emergent phases in the presence of a finite two-body interaction $U$ at finite temperature $T$ and doping $x$. The doping is defined as the deviation of the density from half-filling, i.e. 
\begin{equation}
x=N/N_{sites}-1,
\end{equation}
where $N$ is the total number of particles and $N_{sites}$ is the total number of sites. We give a brief summary of our findings as follows: First, we present the $T$ vs $U$ phase diagram at half-filling, i.e. $x=0$, and $\delta=0$, in figure~\ref{fig2}. For the particle-hole symmetric interaction term (equation \ref{eq_Hu}), $x=0$ is given by the chemical potential $\mu_{A(B/C)}=0$. We show the local magnetization at different sites, $m_{A(B/C)}$, for varying Hubbard interaction at different temperatures $T$ in figure~\ref{fig3}, the lowest temperature being $T=0.01$ for the R-DMFT+CTINT calculations. Due to the presence of the flat band which is distributed over the $B$ and $C$ sites, an infinitesimally small value of $U$ at the zero temperature limit localizes the particles with a sharp increase in the local magnetic moment. At finite temperatures, there is a magnetically ordered metallic phase  discussed in figure~\ref{fig4}. The non-magnetic metallic region at finite temperature displays non-Fermi-liquid (NFL) behavior, which is the concomitant of the flat band singularity, as shown by the non-analytic structure of the local self-energy in figure~\ref{fig5}. We also study the effect of the flat band on the double occupancy $\langle n_\uparrow \ n_\downarrow \rangle$, a direct measure of correlation effects~\cite{Jordens2008}, and present it in figure~\ref{fig6} for varying interactions at different temperatures. We show the effect of the dimerization parameter $\delta$ in figure~\ref{fig7}.
For the doped case $x \neq 0$, we show the emergence of the stripe order with finite wavelengths (see figures~\ref{fig9}-~\ref{fig10}). We also discuss the melting of such stripe order and present a schematic $U$ vs $x$ phase diagram in figure~\ref{fig12}.

\subsection{Finite temperature phase diagram at half-filling}
\label{sec:finitephase}
\begin{figure}[h!] 
\centering
\includegraphics[scale=0.38]{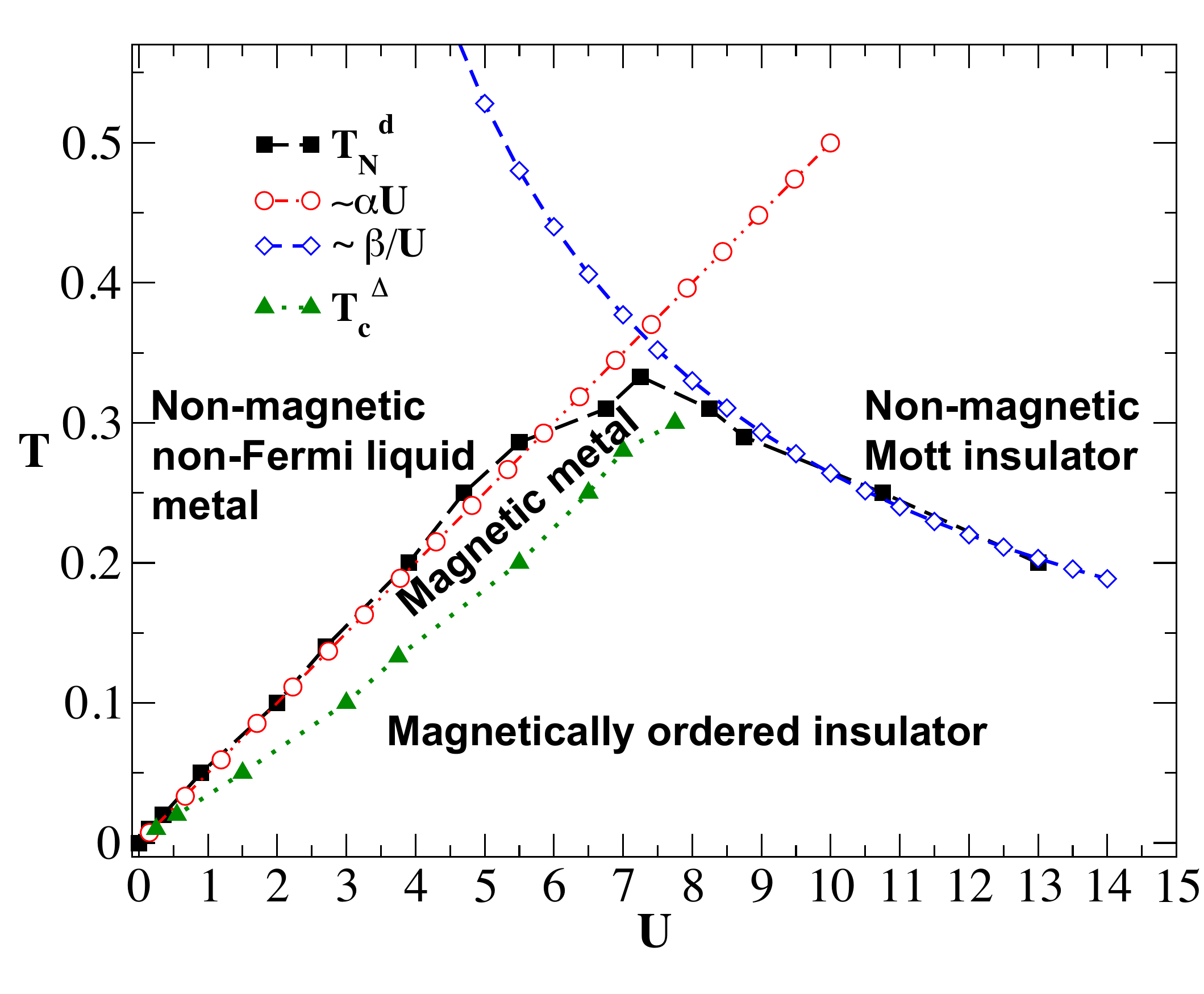}
\includegraphics[scale=0.75]{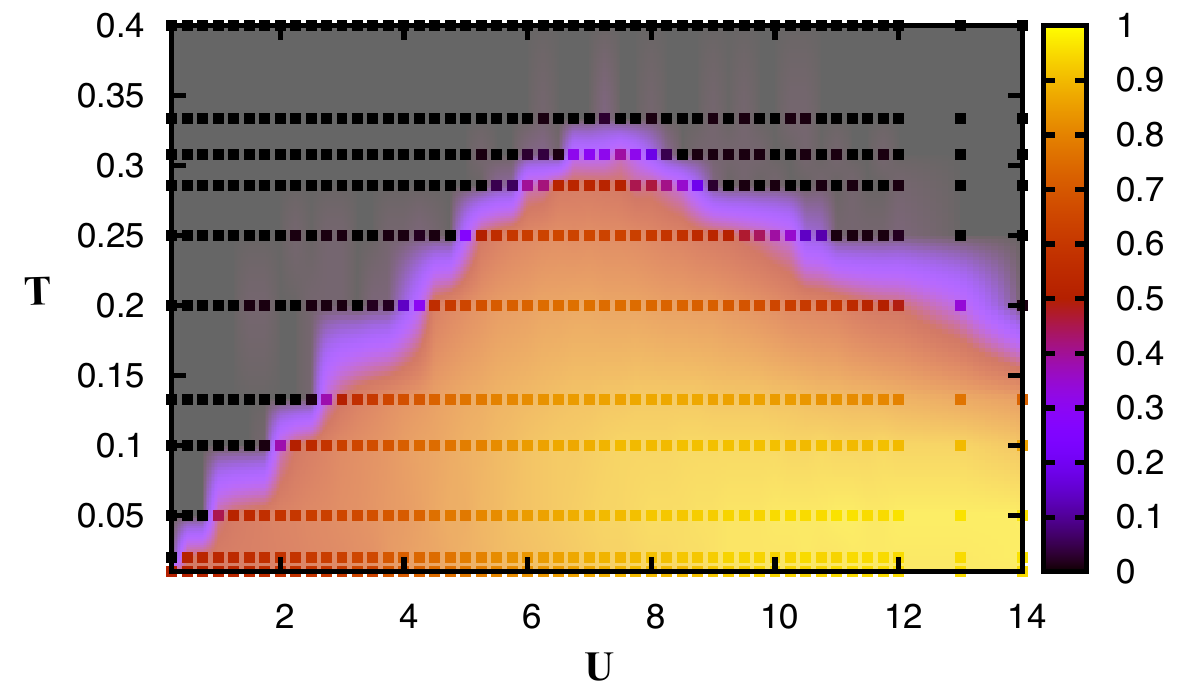}
\caption{Upper panel: Finite temperature phase diagram of the Hubbard model on the Lieb lattice for $x=0$ and $\delta=0$. The filled squares with dashed line represent the N{\'e}el temperature, $T_N^d$, obtained using DMFT~\cite{PhysRevB.95.235109} at interaction strength $U$. The triangles with the dotted lines represent the critical temperature, $T_c^{\Delta}$, at interaction $U$, where a transition from the magnetically ordered metallic state to the magnetically ordered gapped state occurs. The symbol $\Delta$ represent the spectral weight at Fermi-level (metallic behavior) and is defined in the text. The open circles with dashed line represent a linear fit with $\alpha=0.05$. The open diamonds with dashed lines show another fitted curve for the strongly correlated regime where the constant $\beta=2.64$. Lower panel: The color scale corresponds to the magnitude of $m=max(m_A,m_B,m_C)$. The squares are the data points where we have carried out R-DMFT+CTINT calculations.}\label{fig2}
\end{figure}
The finite temperature phase diagram of the repulsive Hubbard model at half-filling, i.e. $x=0$, is shown in Fig.~\ref{fig2}. We allow the breaking of the $SU(2)$ spin rotation symmetry to capture the magnetically ordered phase. In the main panel, we show the variation of $T_N^d$ with $U$. There is a dome-like structure similar to that obtained for the square lattice~\cite{PhysRevB.83.085102,PhysRevB.95.235109}. For $T<max(T_N^d)$ the system traverses to three different regions as the interaction strength is increased. At, for instance, temperature $T=0.20$, there is a non-magnetic NFL metallic region to the left of the dome, a magnetically ordered state within the dome, and a non-magnetic Mott insulating state to the right of the dome. There is a finite region of the phase diagram where we observe a magnetically ordered metal which ultimately gets gapped below $T^\Delta_c$ at a given interaction strength $U$.

In the weakly correlated regime, $T_N^d(U)$ (shown by filled squares with dashed lines) varies linearly as a function of $U$ with $U_c \sim 0$. To show this, we have carried out a linear fit, i.e. $\alpha U$ with $\alpha=0.05$, represented by open circles with a dash-dotted line. In previous studies, a linear behavior of the critical temperature has been predicted by the Bardeen-Cooper-Schrieffer (BCS) theory for the attractive Hubbard model in presence of the flat band~\cite{PhysRevB.83.220503,Heikkilä2011,PhysRevLett.117.045303}. In a recent DMFT study, such linear behavior has also been reported for three dimensional layered Lieb lattice with anisotropic hopping~\cite{PhysRevA.91.063610}. There, for  interlayer hopping $t_z=0.1$, when the flat band contribution is significant, a linear behavior of the ordering temperature with varying interaction strength $U$ has been observed. The onset of the linear behavior occurs at a finite value of $T$ and $U$ due to the finite value of $t_z$. The linear behavior of the ordering temperature with $U$ has also been argued by solving a mean-field gap equation in the presence of the flat band~\cite{PhysRevA.91.063610}. Linear behavior of the pairing in the attractive Lieb lattice Hubbard model, which is equivalent to the zero doping case via the particle-hole transformation has also been shown~\cite{PhysRevLett.117.045303}.

In the strong coupling limit, where particles localize due to strong correlations, the model can be well described by an effective antiferromagnetic Heisenberg model and thus the ordering temperature varies inversely with the interaction, i.e. $T_N^d \propto 1/U$. We present the fitted function $2.64/U$ by open diamonds with a dashed line. In the lower panel of Fig.~\ref{fig2}, we show the magnetic order parameter $m=max(m_A,m_B,m_C)$ as a function of $U$ and $T$ interpolated from the data points marked by the squares.

\begin{figure}[h!]
\centering
\includegraphics[scale=0.62]{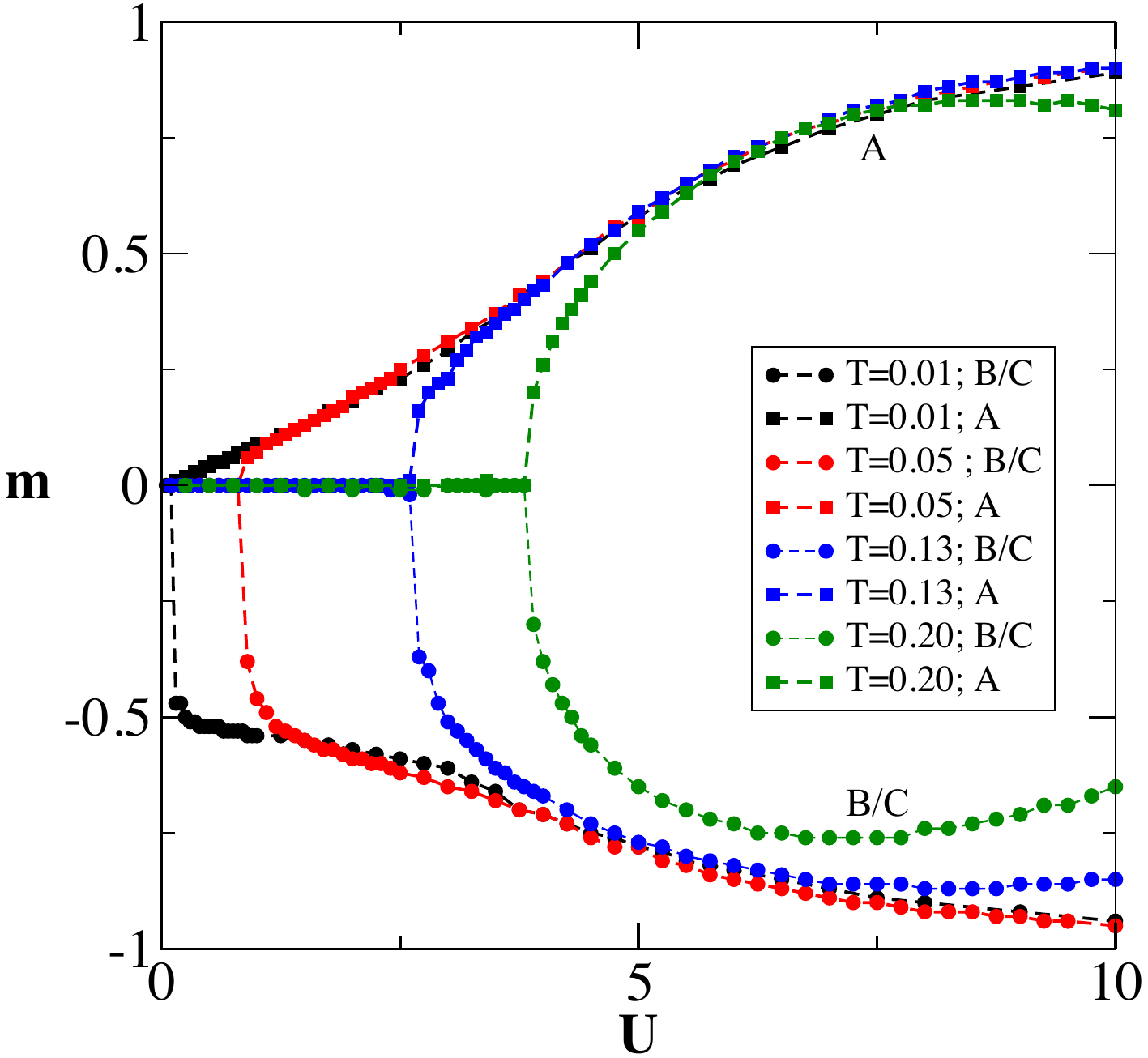}
\caption{Magnetic order $m_{A(B/C)}$ for varying $U$ and different $T$. The critical interaction increases with increasing $T$.}\label{fig3}
\end{figure}

To explore the magnetically ordered phase, we plot $m_{A(B/C)}$ as a function of $U$ at different temperatures $T$ and $\delta=0$ in Fig.~\ref{fig3}. For the smallest temperature $T=0.01$, the magnetic order for the $B$ and $C$ sites changes sharply at $U_c \sim 0.20$ from $0.0$ to $0.5$ while it smoothly assumes a finite value for site $A$. The total magnetization per unit cell, i.e. $m_{tot}=m_A+m_B+m_C\approx-1.0$, is independent of $U$, in accordance with Lieb theorem~\cite{PhysRevLett.62.1201}. Also, the magnetic order behaves linearly with varying $U$ up to $U \sim 4.0$. For a moderate temperature, e.g. $T=0.05$, the critical value of the interaction strength $U_c$ shifts to $~0.90$ and magnetic order assumes a finite value sharply and simultaneously for all the sites. The linear behavior is still visible in $m_{A(B/C)}(U)$. There is smooth crossover from flat band ferromagnetic behavior to strong coupling  Heisenberg ferrimagnetic behavior with increasing interaction $U$. For high temperatures, such as the case $T=0.13$ and beyond, the linear behavior is no more visible and the net magnetization per unit cell is $U$ dependent and thus the Lieb theorem is no more satisfied. The enhance magnetism for $U\sim 0$, even when $\delta=0$, is the consequence of present flat band in the dispersion and the contribution of the other bands are negligible~\cite{PhysRevLett.117.045303}. For system with semi-metal dispersion, e.g. honeycomb lattice, there is associated finite critical value of Hubbard interaction at which semi-metal to antiferromagnetic transition occurs~\cite{PhysRevB.90.085146}, in contrast the system gets magnetized immediately with the onset of $U$ for the Lieb lattice. The emergent physics appears to be mainly driven by the presence of the flat-band at small interactions.
\begin{figure}[h!]
\centering
\includegraphics[scale=0.55]{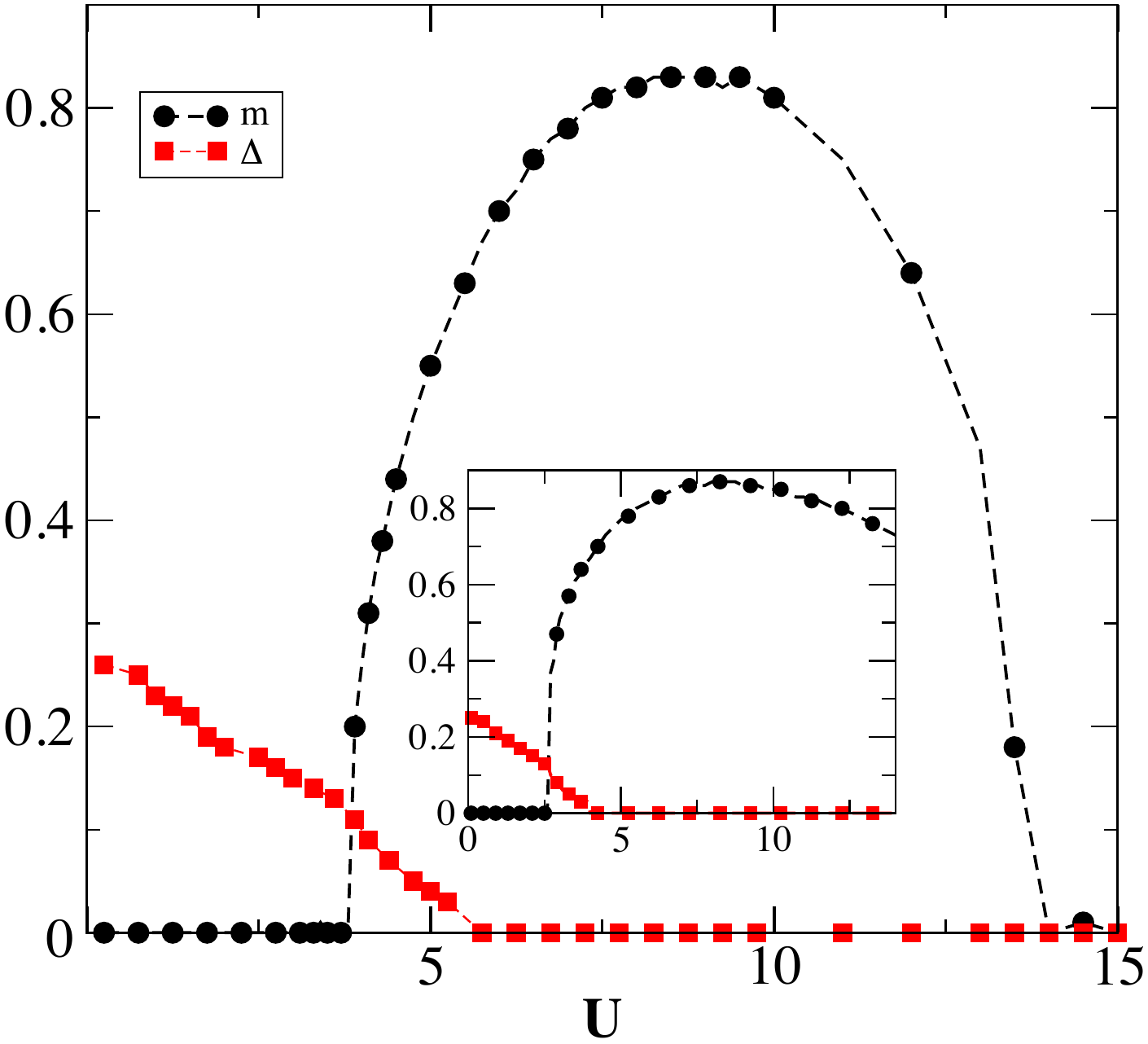}
\caption{ In the main panel: The magnetic order parameter $m=max(m_A,m_B,m_C)$ and the spectral weight at Fermi-level $\Delta=max(\Delta_A,\Delta_B,\Delta_C)$ varying with $U$ for $T=0.20$. There is a finite region where the magnetic phase has a finite $\Delta$, signifying the metallic behavior. In the inset: Similar behavior has been presented for $T=0.13$. }\label{fig4}
\end{figure}

To understand the magnetically ordered metallic region, we show the magnetic order, i.e. $m=max(m_A,m_B,m_C)$, and a measure of spectral weight at Fermi-level, i.e. $\Delta=max(\Delta_A,\Delta_B,\Delta_C)$ in Fig.~\ref{fig4}. $\Delta_{A(B/C)}=-G_{A(B/C)}(\tau=1/2T)$~\cite{0295-5075-84-3-37009} (where $G_{A(B/C)}(\tau)$ is the local imaginary time Green's function) is zero for a gapped system, while it assumes a finite value for a gapless system. In the main panel, we show $\Delta$ and $m$ for $T=0.20$. There is a finite region of $U$ between $3.75-5.75$, where both quantities assume finite value, showing the existence of magnetically ordered metallic region. In the inset, we show a similar analysis for $T=0.13$, where the region is still finite but smaller than for $T=0.20$. The magnetically ordered metallic region gets narrower with decreasing temperature as show in Fig.~\ref{fig2}.

\subsection{Non-Fermi liquid behavior}

We explore the finite-temperature quasi-particle behavior in the normal state in the weak coupling regime in the presence of a flat band. We find the breakdown of the usual Fermi-liquid behavior in the region by observing the scattering rate, i.e. the imaginary part of the local self-energy, for the different sites within the unit cell. There have been a few studies using perturbation theory and a renormalization group approach predicting non-Fermi liquid behavior due to the presence of, for example, Van Hove or power law singularities in the dispersion of the noninteracting part of the Hamiltonian~\cite{PhysRevB.46.11798,PhysRevB.51.9253,refId0,PhysRevB.68.195101}. Phenomenological marginal Fermi-liquid~\cite{PhysRevLett.63.1996} behaviour, where the self-energy has a linear frequency dependence, has been proposed in the context of the cuprates. A diverging non-interacting density of states leads to a soft-gap in the effective hybridization function of DMFT and consequently to a non-Fermi-liquid signature in the local self-energy~\cite{PhysRevB.80.245112,PhysRevB.82.155126}. Non-Fermi-liquid behaviour has also been studied using theories which include non-local correlations~\cite{PhysRevB.79.045133,PhysRevB.80.165126}. For a well defined Fermi liquid, the self-energy for low Matsubara frequencies $\omega_n$ can be written as
\begin{equation}
\Sigma(i\omega_n)=a \ i\omega_n  +b\label{eq9}
\end{equation}
where $a$ and $b$ are real constants. The quasi-particle weight $Z=m/m*$, where $m$ is the bare mass and $m*$ is the mass in the presence of many-body effects, can be defined in terms of the self-energy as
\begin{equation}
Z=\big(1-\frac{\partial \text{Im}\Sigma(i\omega_n)}{\partial \omega_n}|_{n\rightarrow 0}\big)\label{eq10}
\end{equation} 
\begin{figure}[h!]
\centering
\includegraphics[scale=0.55]{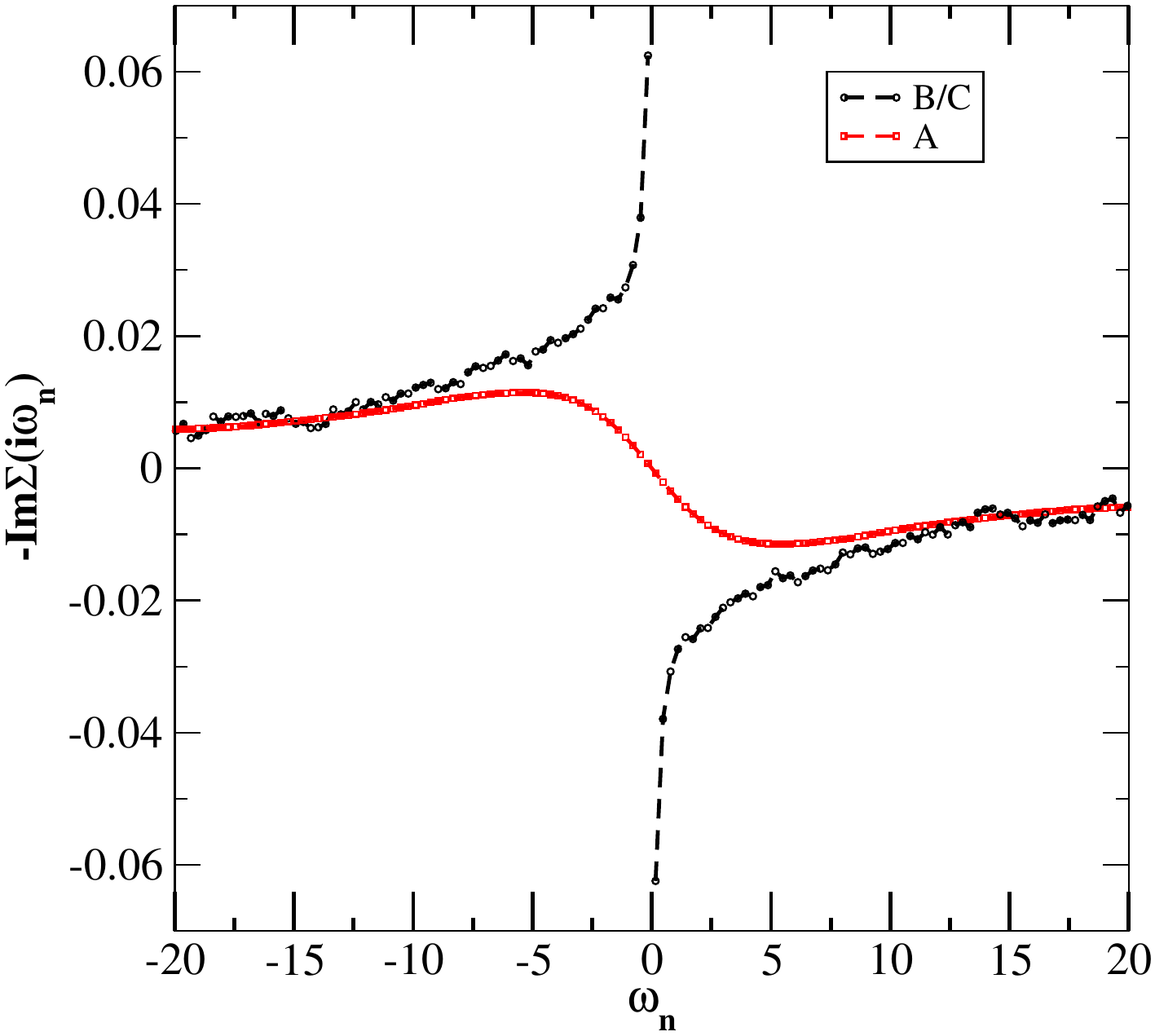}
\includegraphics[scale=0.55]{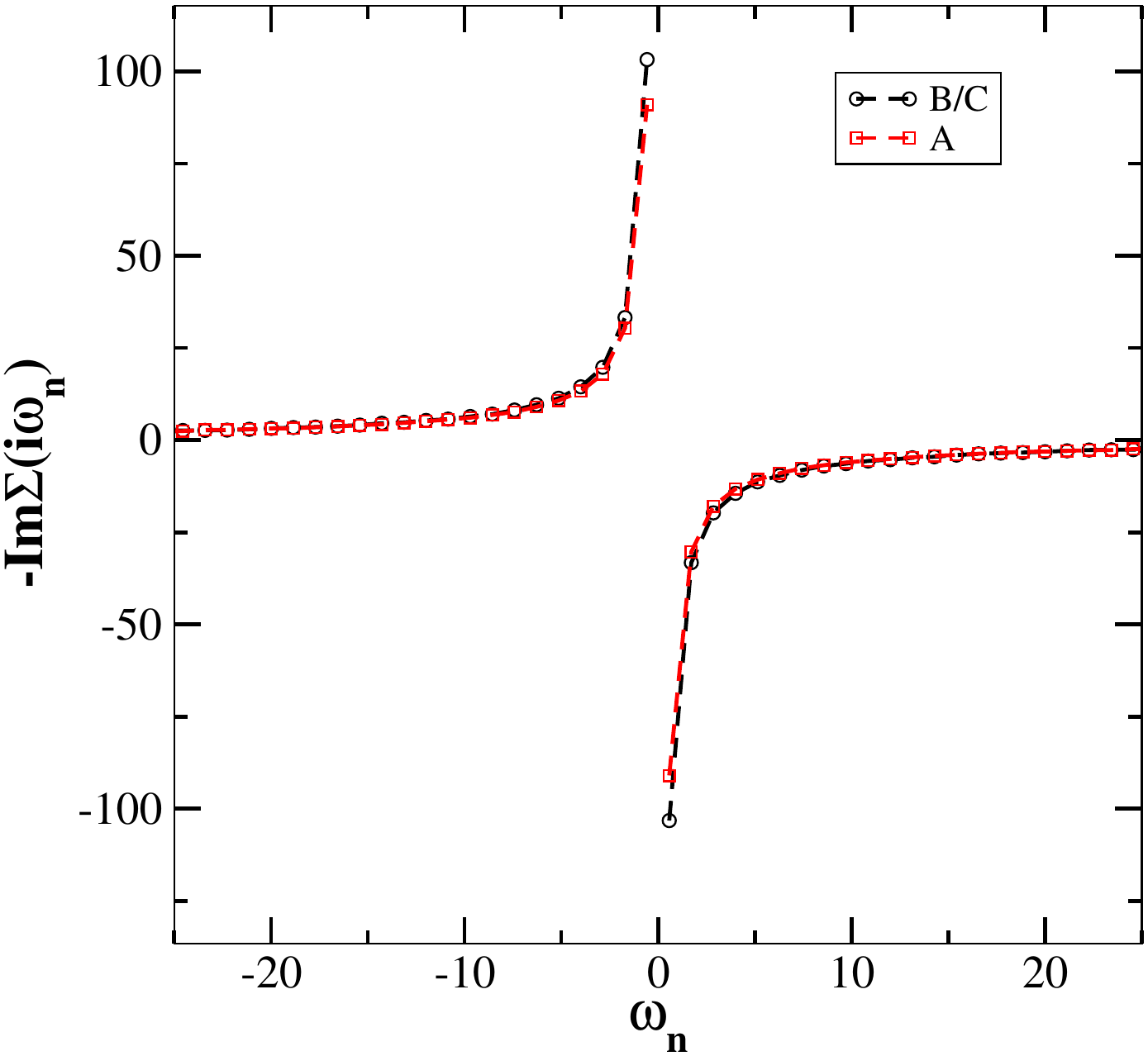}
\caption{In the upper panel: Imaginary part of the local self-energy, i.e. $-\text{Im}\Sigma(i\omega_n)$ vs Matsubara frequency $\omega_n$, for different sites $A$, $B$ and $C$, for $U=0.70$ and $T=0.05$. For these parameters the system is in the non-magnetic metallic regime (see Fig.~\ref{fig2}). In the lower panel: $-\text{Im}\Sigma(i\omega_n)$ for the different sites vs $\omega_n$ for $U=16$ and $T=0.18$ when system is in the Mott insulating regime. Here the dimerzation parameter $\delta=0$, but we have observed the same quantitative behavior also for $\delta\neq 0$.}\label{fig5}
\end{figure}
We observe the imaginary part of the self-energy at the lowest numerically calculated Matsubara frequency $\omega_0$ and at the next consecutive frequency $\omega_1$. For the Fermi liquid behavior $|\text{Im}(i\omega_0)| < |\text{Im}(i\omega_1)| $ while for the non-Fermi liquids $|\text{Im}(i\omega_0)| > |\text{Im}(i\omega_1)| $. Additionally, the scattering rate~\cite{PhysRevLett.101.146403} per unit cell, an estimate of the conductivity, can be given as
\begin{equation}
\tau^{-1}=\tau_A^{-1}+\tau_B^{-1}+\tau_C^{-1}, \label{eq11}
\end{equation}
where $\tau_{A(B/C)}^{-1}=-\text{\Im}\Sigma(i\omega_n=0)$ is diverging and thus violating the Fermi-liquid behavior. In the upper panel of Fig.~\ref{fig5}, we show the imaginary part of the self-energy in the non-magnetic region of the phase diagram shown in Fig.~\ref{fig2}. The self-energy for the $B(C)$ site, which carry the flat band, diverges for small frequencies $|\omega_n|$, while the self-energy for the site $A$ is still analytical. In  the lower panel, we show the self-energy for the Mott insulating regimes of the phase diagram. The self-energies for all sites, i.e. $A$, $B$ and $C$, are diverging for $\omega_n\rightarrow 0$, a key feature of Mott insulators. Non-Fermi liquid behaviour in the presence of a flat band has been discussed for a multiband lattice Hamiltonian in the presence of an attractive Hubbard interaction using perturbation theory~\cite{PhysRevB.94.245149}. We conclude that the presence of the flat band, causing singular behavior at the Fermi level, leads to the NFL behavior at finite temperature in the non-magnetic weakly interacting regime. Such NFL behavior present at weak coupling will be missed within static mean field theories, where the dynamical part of the self-energy is zero. To further explore the interaction effects in the presence of the flat band, we additionally present the double occupancy behavior with $U$ and $T$.

\subsection{Double occupancy}
\label{subsec:double}
\begin{figure}[h!]
\centering
\includegraphics[scale=0.6]{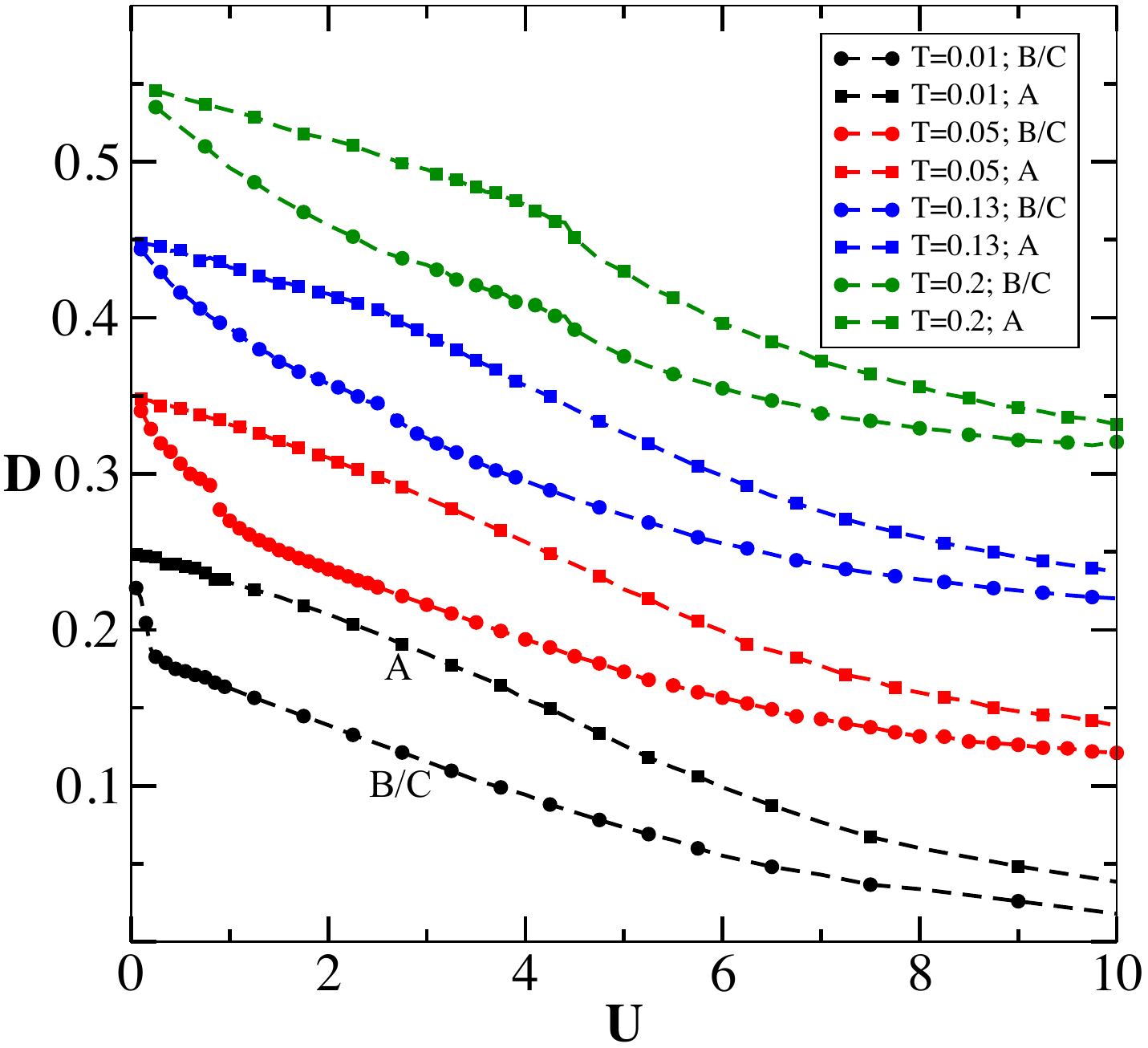}
\caption{Double occupancy, i.e. $D=\langle n_{\uparrow} n_{\downarrow}\rangle$, for the different sites $A$, $B$ and $C$ vs $U$ for different $T$. For the purpose of clarity, we have added an offset in $y$- axis. The size of the offset is $0.0, 0.1, 0.2$ and $0.3$ for $T=0.01, 0.05, 0.13$ and $0.20$ respectively.}\label{fig6}
\end{figure}

The double occupancy $D$ represents the probability of two particles to occupy the same site. It is $0.25$ in the zero interaction limit while it vanishes in the Mott insulating large $U$ limit. Double occupancy at a given site with the DMFT+CTINT solver can be evaluated from the Monte-Carlo perturbation order~\cite{PhysRevB.76.035116} given by
\begin{equation}
\langle k \rangle_{\text{MC}}=-\beta U \Bigg\langle \Bigg(n_{i\uparrow}-\frac{1}{2}\bigg)\Bigg(n_{i\downarrow}-\frac{1}{2}\bigg)-\epsilon^2\bigg\rangle
\end{equation}
where $\epsilon$ is the impurity solver parameter chosen to be small for the half-filled case. Further, the double occupancy can be given as
\begin{eqnarray}
D=\langle n_{i\uparrow} n_{i\downarrow} \rangle= \frac{n}{2}-\frac{\langle k \rangle_{\text{MC}}}{\beta U}-\frac{1}{4}+\epsilon^2, \label{eq12}
\end{eqnarray}
where $n=n_{\uparrow}+n_{\downarrow}=1$ for half-filling. In Fig~\ref{fig6}, we show the double occupancy varying with increasing $U$ for different sites in the unit cell at different temperatures. For weak interactions and low temperatures, the double occupancy decreases smoothly for site $A$ while it changes sharply for the $B$ and $C$ sites with a kink at the transition point. Double occupancy is smaller for the $B$ and $C$ sites that carry the flat band compared to the $A$ site for a given interaction $U$ and temperature $T$. For Large temperatures, the kink is visible in double occupancy for all the sites. In the presence of the flat band, even infinitesimal  interaction favors enhanced localization of the particles demonstrated by the sharp change of the double occupancy.  In the strongly interacting limit the double occupancy for all the sites $A$, $B$ and $C$ coalesces and vanishes.

The double occupancy for a site can directly be compared with the local moment $m_z^2$ measured in the experiments~\cite{Jordens2008}, given as
\begin{equation}
\langle m_z^2 \rangle =1-2\langle n_{i\uparrow}n_{i\downarrow}\rangle\label{eq13}.
\end{equation}
Recently, using finite size determinant quantum Monte-Carlo, $\langle m_z^2 \rangle$ with varying $U$ for the Lieb lattice has been reported~\cite{PhysRevB.94.155107}, with results consistent with our findings from R-DMFT+CTINT calculations.
\subsection{Tuning the flat band contribution}
\label{sec:dimer}
\begin{figure}[h!]
\centering
\includegraphics[scale=0.6]{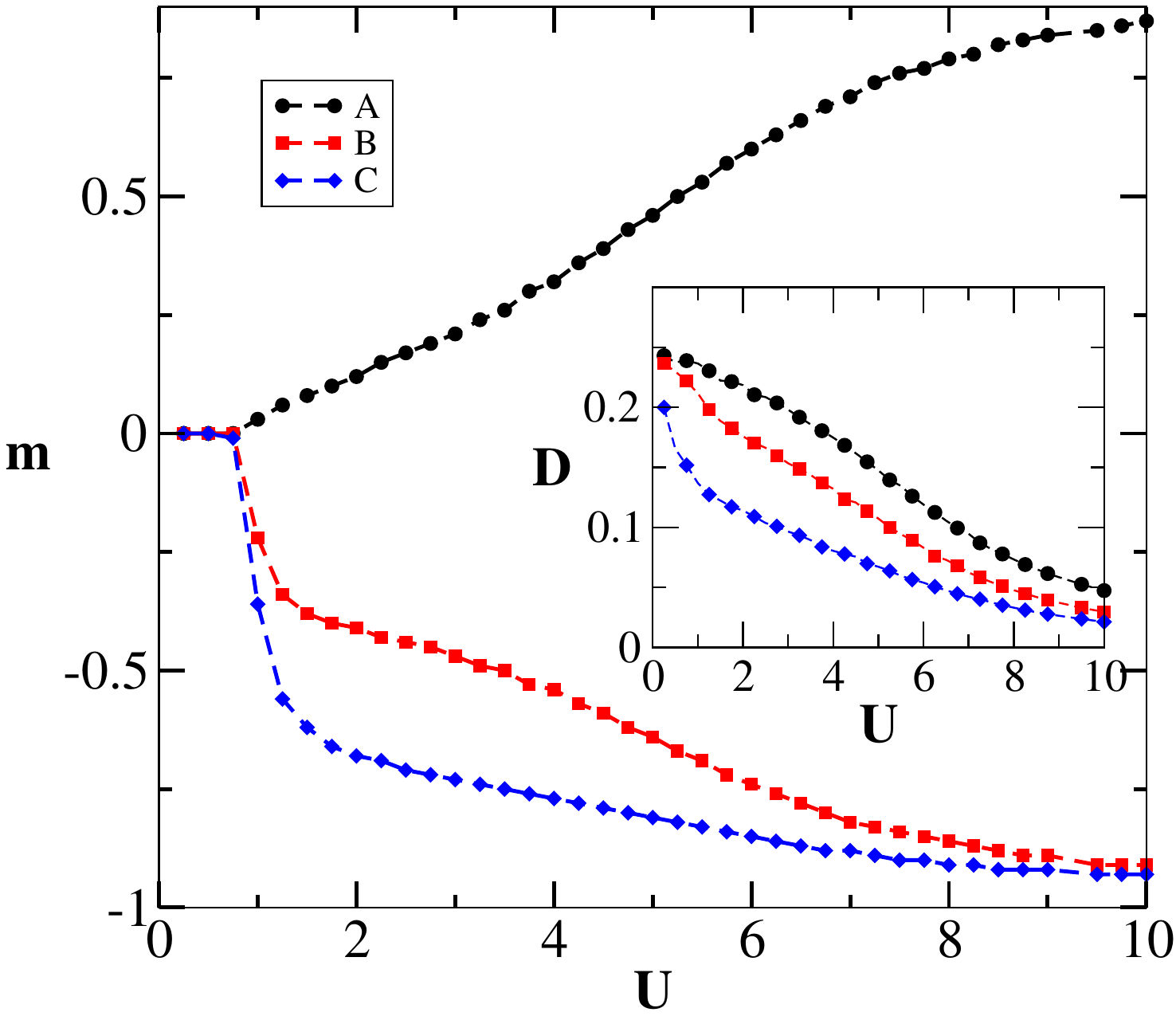}
\caption{In the main panel: Magnetic order for $A$, $B$ and $C$ sites for varying $U$ at $T=0.05$. The dimerization parameter is $\delta=0.6$. In the inset: The double occupancy for the same parameters as in the main panel.}\label{fig7}
\end{figure}

As discussed in section~\ref{sec:model}, the contribution of the flat band at different sites in the unit cell can be tuned by varying the dimerization parameter $\delta$. We show the magnetic ordering $m_{A(B/C)}$ for the site $A(B/C)$ for varying interaction strength $U$ at temperature $T=0.05$ for finite $\delta$ in the main panel of Fig.~\ref{fig7}. For a moderate value of $U$, $|m_C|>|m_B|$ showing that the $C$ site has more weight of the flat band than the $B$ site, unlike in the $\delta=0.0$ case discussed in section \ref{subsec:double} where the $B$ and $C$ sites are equivalent. The $m_B$ and $m_C$ tend to the same value in the large $U$ limit, where the flat band behavior crosses over to a strong coupling behavior. The trend in the double occupancy in presence of the dimerization is similar to the the magnetic order as shown in the inset of Fig.~\ref{fig7}.

We conclude that introducing such partial dimerization in the hopping can be used as a tool to infer the contribution of the flat band to different spatially resolved quantities.

\subsection{Doping induced stripe order}

\begin{figure}[h!]
\centering
\includegraphics[scale=0.43]{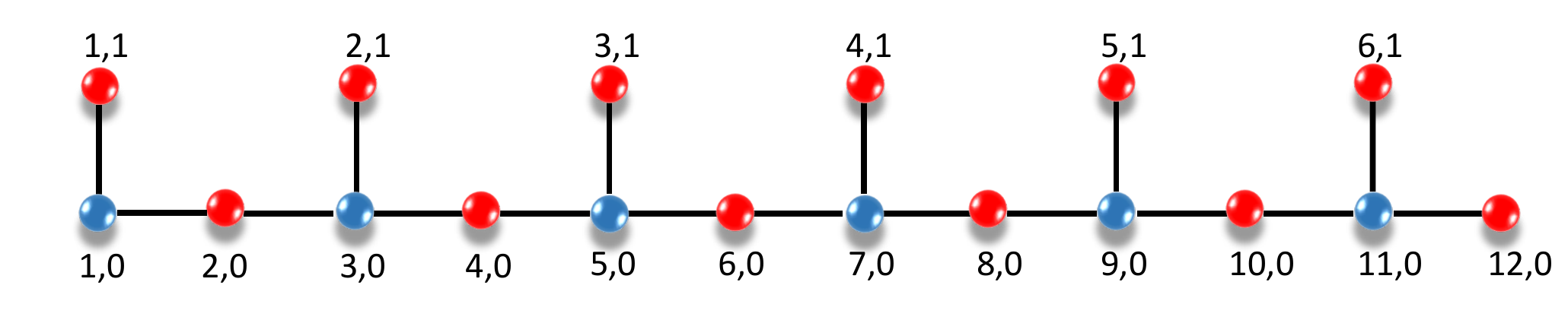}
\caption{Schematic diagram of the unit cell chosen for studying the stripe order.}\label{fig8}
\end{figure}

To explore the possible stripe order for the Lieb lattice away from half-filling, we have carried out R-DMFT+CTINT calculations using unit cells with a maximum of $36$ sites. In Fig.~\ref{fig8}, we show a schematic diagram of the unit cell with $18$ sites. Real space positions of the sites in the unit cell are labeled by the indices $(r_x, r_y)$. Sites of the same color have equivalent order parameters at half-filling and zero dimerization.
We uniformly dope the system by choosing a finite chemical potential $\mu(r_x, r_y)$ independent of $r_x$ and $r_y$, and observe the emergent stripe order, which simultaneously displays spin density wave and charge density wave order. The doping $x$ is defined as
\begin{equation}
x=\sum_{r_x, r_y}\frac{n(r_x, r_y)}{N}-1
\end{equation}
where $n(r_x,r_y)=n(r_x,r_y,\uparrow)+n(r_x,r_y,\downarrow)$ is the density of site with index $(r_x,r_y)$ and $N$ is the total number of sites in the unit cell. We also explore the effect of temperature on such stripe order.

\begin{figure}[h!]
\centering
\includegraphics[scale=0.55]{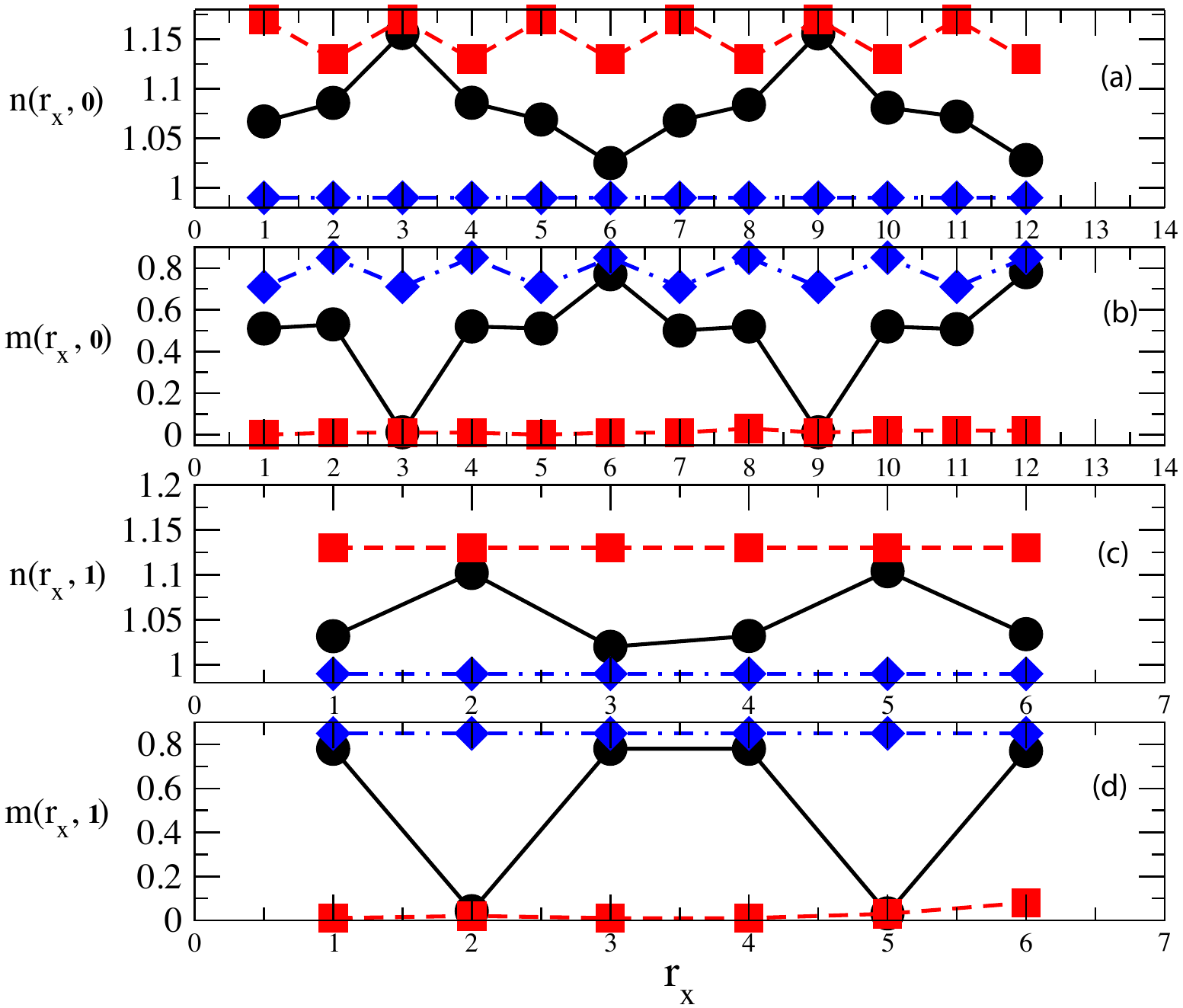}
\caption{In panel (a): $n(r_x,0)$ (Eq.~\ref{eq8}) vs $r_x$.  Blue diamonds with a dash-dotted line, black solid circles with a solid line and red squares with a dashed line correspond to $x=0.0, 0.07$ and $x=0.14$ respectively. In panel (b): $m(r_x,0)$ (Eq.~\ref{eq7}) vs $r_x$. Colors and symbols are in accordance with (a). In panel (c): $n(r_x,1)$ vs $r_x$. In panel (d): $m(r_x,1)$ vs $r_x$. The value of the temperature is $T=0.05$ and interaction is $U=6.0$.}\label{fig9}
\end{figure}

We show $n(r_x, 0)$ and $n(r_x, 1)$ for different sites of the unit cell for the zero doping case in Fig.~\ref{fig9}(a) and Fig.~\ref{fig9}(c), respectively. Here the density is uniform with $n(r_x, r_y)=1$ for all $r_x$. The variation of $m(r_x,0)$ and $m(r_x,1)$ (see equation \ref{eq7}) has been presented in Fig.~\ref{fig9}(b) and Fig.~\ref{fig9}(d), respectively. The $m(r_x,0)$ has a sub-lattice ordering while the $m(r_x,1)$ is constant for all $r_x$, consistent with the bipartite structure of the Lieb lattice.

For moderate doping $x=0.07$, there is a charge density wave (CDW) with a finite wavelength shown by $n(r_x, 0)$ and $n(r_x, 1)$ in Fig.~\ref{fig9}(a) and Fig.~\ref{fig9}(c), respectively. Similarly, a spin density wave (SDW) emerges with the wavelength of $12$ sites as presented in Fig.~\ref{fig9}(b) and Fig.~\ref{fig9}(d) by the behavior of $m(r_x,0)$ and $m(r_x,1)$, respectively. This is a so-called vertical stripe state, where the simultaneous SDW and CDW are directed along the bonds of the lattice (as opposed to e.g. diagonally). We have also carried out R-DMFT+CTINT calculations with a larger number of sites in the unit cell by doubling the size to $36$ sites. The stripe order is stable for the larger unit cell as well.

Increasing the doping further to $x=0.14$, the finite wavelength charge order turns into a sub-lattice ordering where the $A$ sites have a different density than the $B$ and $C$ sites, but the translational and rotational symmetries of the lattice are not broken. This has been shown in Fig.~\ref{fig9}(a) and Fig.~\ref{fig9}(c). The magnetic ordering $m(r_x,r_y)$ vanishes for all sites as visible in Fig.~\ref{fig9}(b) and Fig.~\ref{fig9}(d). The decrease in wavelength with increasing doping is consistent with mean-field findings~\cite{0953-8984-23-50-505601} and has been reported for high $T_{c}$ superconductors~\cite{PhysRevB.79.100502}. The increase in the wavevector (decrease in wavelength) can also be argued from the FFLO state appearing for doped attractive Hubbard model, which can be related to the stripe order. The increasing doping corresponds to increased imbalance in the Fermi-surface mismatch of the two components and thus a large wave vector is required for pairing to be possible~\cite{kinnunen2017fulde}.

\begin{figure}[h!]
\centering
\includegraphics[scale=0.55]{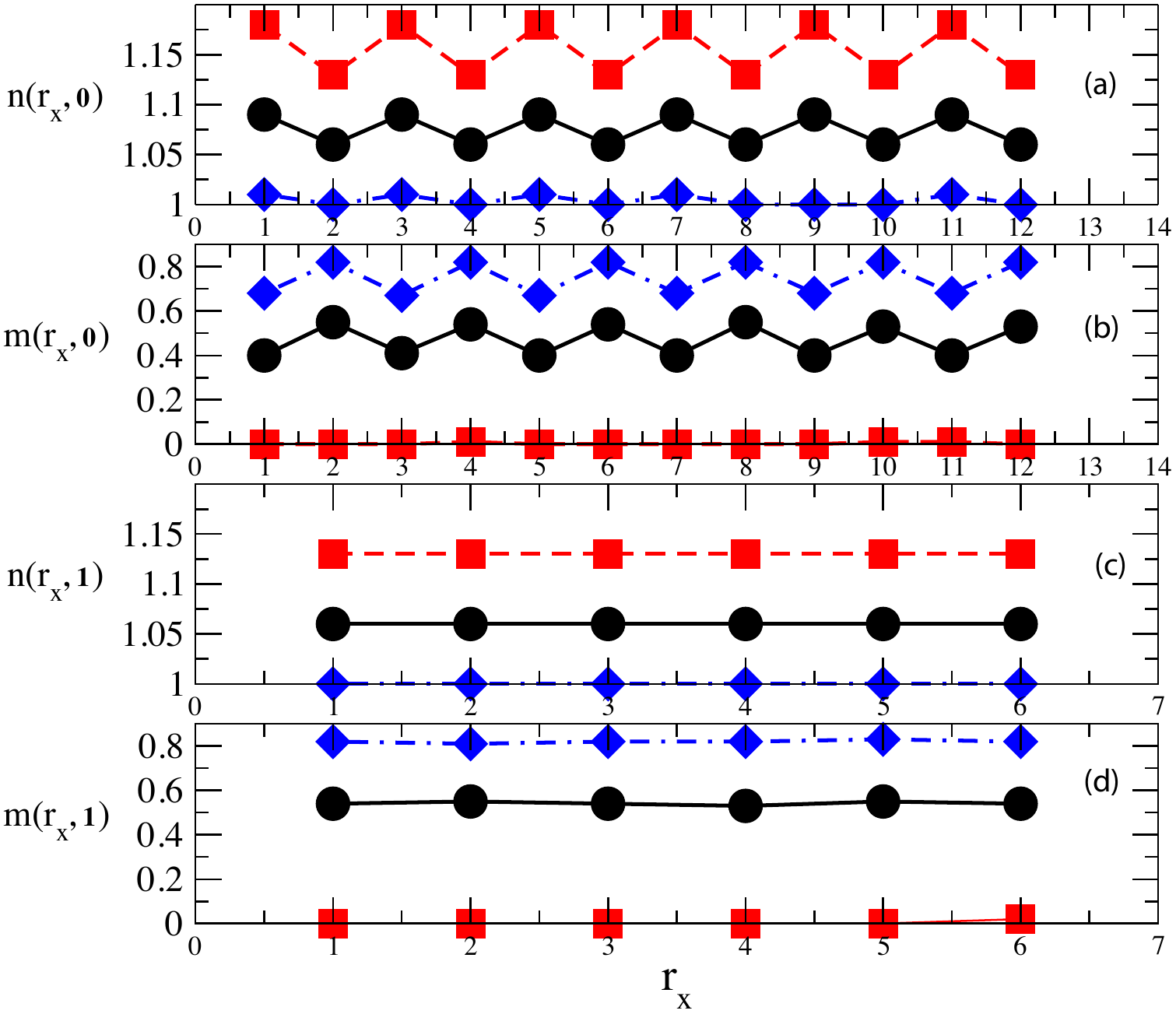}
\caption{The density $n(r_x,r_y)$ and magnetic ordering $m(r_x,r_y)$ (In panel (a) and (b): $r_y=0$. In panel (c) and (d): $r_y=1$) has been shown for different doping for $T=0.10$ and $U=6.0$. Blue diamonds with a dash-dotted line, black solid circles with a solid line and red squares with a dashed line correspond to $x=0.0, 0.07$ and $x=0.14$ respectively.}\label{fig10}
\end{figure}

\begin{figure}[h!]
\centering
\includegraphics[scale=0.6]{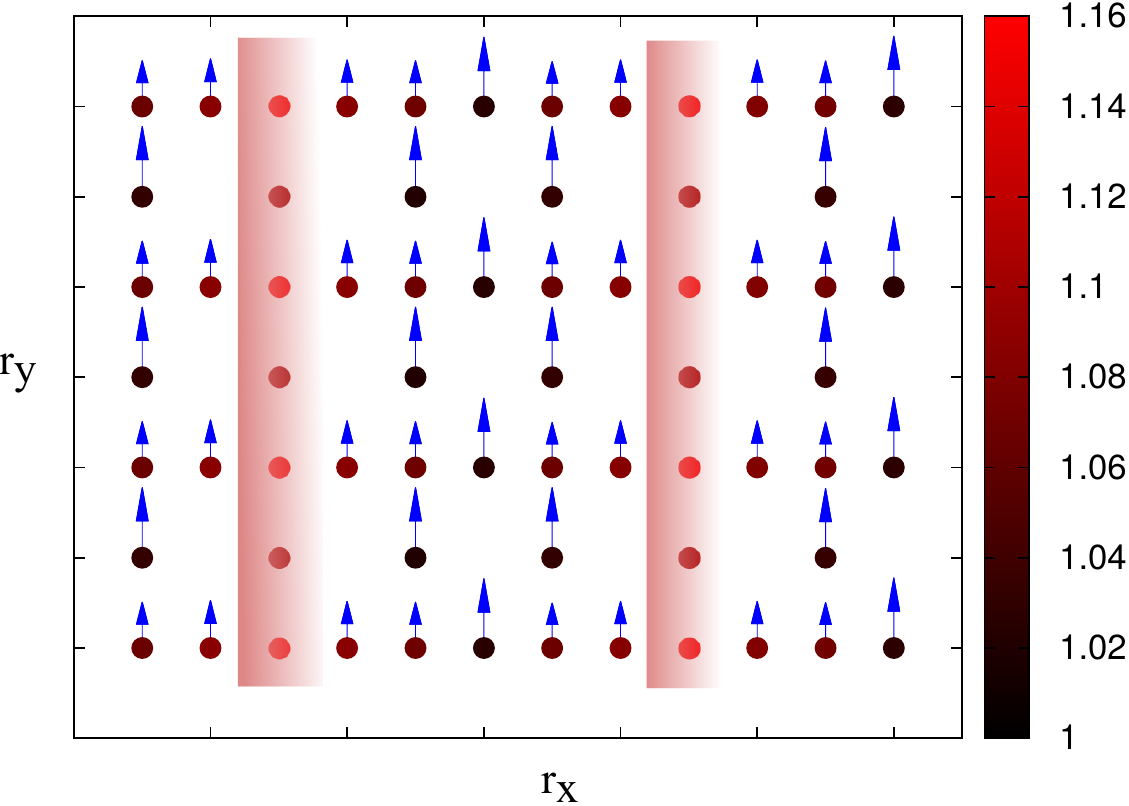}
\includegraphics[scale=0.6]{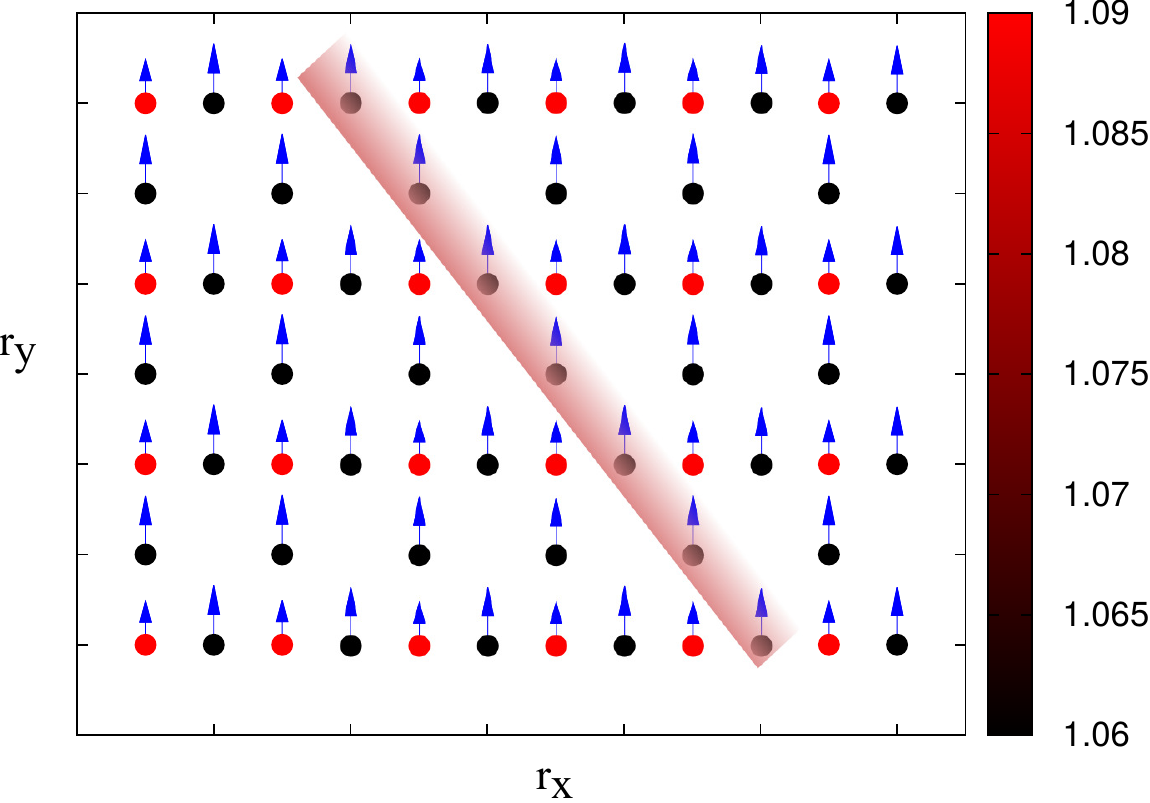}
\caption{Upper panel: the magnetic ordering $m(r_x,r_y)$ and density $n(r_x,r_y)$ for different $(r_x,r_y)$ for $x=0.07$ and $T=0.05$. Size of the arrows represents the magnitude of $m(r_x,r_y)$, while the color of the circles represents the magnitude of $n(r_x,r_y)$ at $(r_x,r_y)$ . In the shaded region the density is maximal with vanishing magnetic order displaying the vertical sripe ordering. Lower panel: $m(r_x,r_y)$ and $n(r_x,r_y)$ for different $(r_x,r_y)$ for $x=0.07$ and $T=0.10$ showing sublattice ordering. The magnitude of the charge and spin orders is constant along the diagonal.}\label{fig11}
\end{figure}

To explore finite temperature effects on the stripe order, we study the doped system for the increased temperature $T=0.10$. The symbols and colors are according to Fig.~\ref{fig9}. In Fig.~\ref{fig10}(a) and Fig.~\ref{fig10}(c) we show the densities $n(r_x,0)$ and $n(r_x,1)$ at half-filling ($x=0.0$), where they are uniform and equal to $1.0$ for all $r_x$, similarly to the $T=0.05$ case. Increasing the doping to $x=0.07$, a sub-lattice ordering emerges in the densities $n(r_x,0)$ and $n(r_x,1)$. This structure prevails for $m(r_x,0)$ and $m(r_x,1)$ as well, as shown in Fig.~\ref{fig10}(b) and Fig.~\ref{fig10}(d), respectively. The simultaneous sublattice ordering in $m(r_x,r_y)$ and $n(r_x,r_y)$ can also be well seen in the lower panel of figure~\ref{fig11}. This state resembles the diagonal stripe order \cite{doi:10.1063/1.4818402,PhysRevLett.113.046402}, where the direction of the stripes is at an angle to the lattice bonds.

For the large doping $x=0.14$, the sub-lattice ordering in the density survives as shown in Fig.~\ref{fig10}(a) and Fig.~\ref{fig10}(c), while the local magnetization $m(r_x,r_y)$ vanishes which can be seen in Fig.~\ref{fig10}(b) and Fig.~\ref{fig10}(d).  

 An important finding of the present work is the presence of the charge order without spin ordering at higher $x$ in contrast to the square lattice where charge and spin order melt simultaneously within R-DMFT~\cite{PhysRevB.82.233101}. Findings of R-DMFT for the square lattice contradict the experimental data on the high-T$_c$ superconductors showing  charge ordering for a wide temperature range with no magnetic ordering~\cite{PhysRevB.79.100502} and the inconsistency~\cite{PhysRevB.82.233101} is attributed to the absence of non-local correlations inherent to the R-DMFT approach. In contrast, the sublattice ordering in the charge sector of the doped Lieb lattice originates from the inequivalent sites in the unit cell rather than from intersite correlations, which  can be captured within the R-DMFT approach. It also provides further evidence that the unveiled stripe order is a robust property of the Lieb lattice.

 To visualize the difference between vertical and diagonal like stripe order, we present a two dimensional spin, i.e. $m(r_x,r_y)$, and charge, i.e. $n(r_x,r_y)$, distribution on the Lieb lattice for $x=0.07$, $T=0.05$ and $x=0.07$, $T=0.10$ in figure~\ref{fig11}(a) and figure~\ref{fig11}(b), respectively. We stack the unit cell shown in figure~\ref{fig8} in the $y$- direction. It is important to mention here that the actual R-DMFT calculation has only been done for that $18$-site unit cell mentioned in figure~\ref{fig8}.
We also note that the repulsive Hubbard model can be mapped to an attractive Hubbard model with a single spin channel particle-hole transformation, i.e.
\begin{eqnarray}
c_{i\downarrow}\longleftrightarrow \epsilon(i) c_{i\downarrow}^\dagger \nonumber \\
c_{i\uparrow}\longleftrightarrow c_{i\uparrow}^\dagger,
\end{eqnarray}
where $\epsilon(i)=1$ for one sublattice of the bipartite lattice and $\epsilon(i)=-1$ for the other . Also, different order parameters for the two cases can be connected~\cite{PhysRevA.79.033620}. For example, the stripe order for the doped repulsive Hubbard model can be connected to the Fulde-Ferrell-Larkin-Ovchinnikov (FFLO) state of the doped attractive $U$ Hubbard model with a finite spin imbalance $\mu_{\uparrow}\neq \mu_{\downarrow}$~\cite{PhysRevLett.98.216402,kinnunen2017fulde}. Therefore, stripe order observed for repulsive Hubbard model on the Lieb lattice predicts the presence of FFLO state in the attractive regime.

\begin{figure}[h!]
\centering
\includegraphics[scale=0.50]{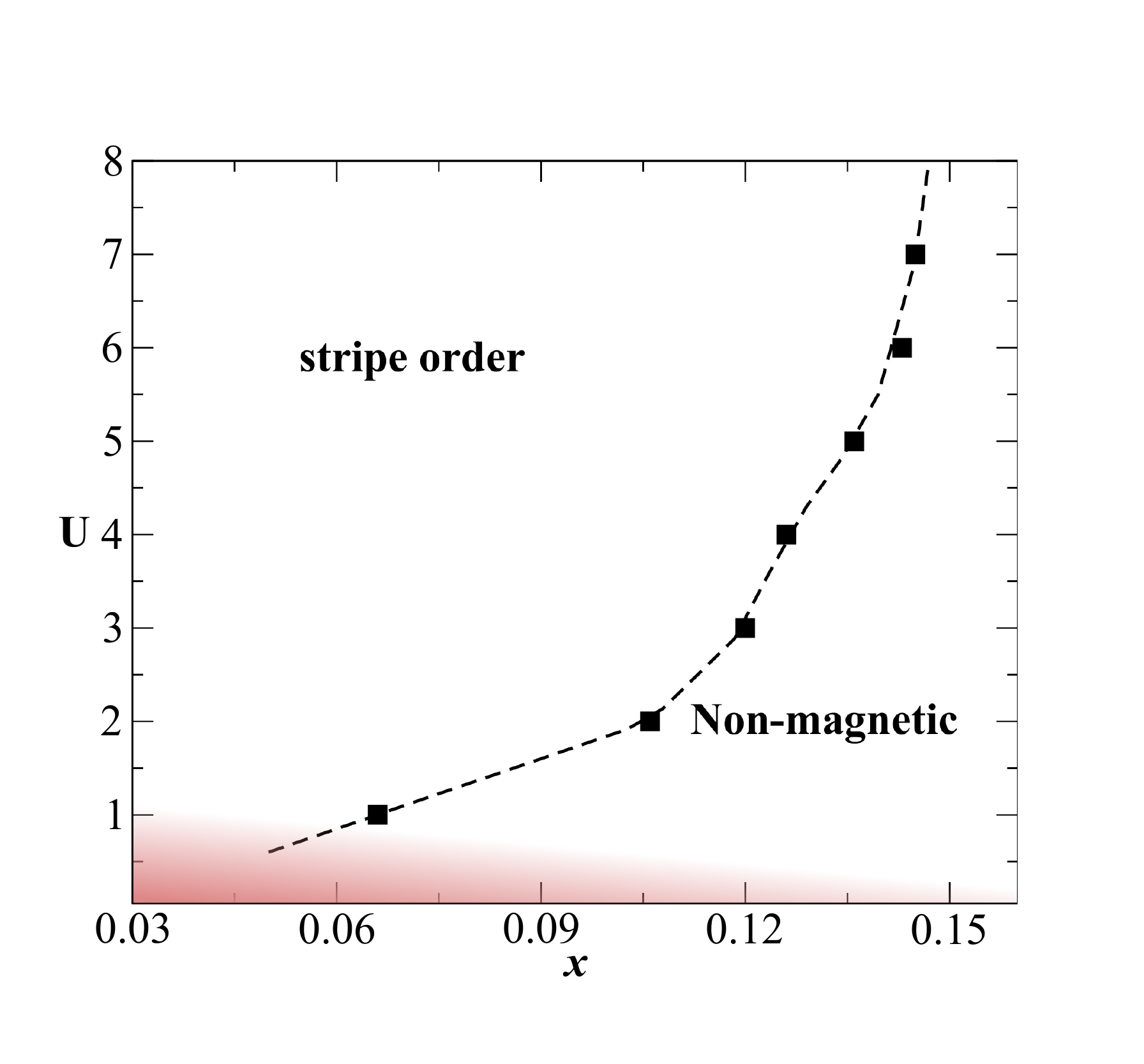}
\caption{Phase diagram for the Hubbard model on the Lieb lattice away from half-filling and $T=0.05$. Filled squares are the transition point obtained using R-DMFT+CTINT calculation. Dashed line is a guide to the eye.}\label{fig12} 
\end{figure}
We summarize our calculation in the phase diagram of the Hubbard model on 2D Lieb lattice obtained by varying doping, i.e. $0.03<x<0.16$, and the interactions shown in Fig.~\ref{fig12} at a fix temperature $T=0.05$. We have carried out R-DMFT+CTINT calculations using unit cells with $36$ sites. For the small interactions, for instance $U=1.0$, we find a stripe order with sub-lattice ordering similar to the one appearing at $T=0.10$ and $U=6.0$ (see Fig.~\ref{fig10}) for $x < 0.06$, while the system is non-magnetic for $x > 0.06$. For large $U$, stripe order with a finite wavelength has been observed. With the increasing interaction strength $U$, the critical value of the doping, i.e. $x_c$, for the transition from stripe order to the non-magnetic order increases and saturates to $x_c \sim 0.15$ for large $U$. For $U<1.0$ and finite doping, shown by the shaded region, the DMFT calculation did not converge with good accuracy. One of the possible reason could be the sharp change in the magnetic order parameter for the given interaction and temperature at half-filling (Fig.~\ref{fig3}). 

\section{Summary and outlook}

We have applied the R-DMFT combined with a CTINT impurity solver to elucidate the influence of flat band on various emergent phases of the repulsive Hubbard model on the $2$D Lieb lattice. At half-filling, we present a full finite temperature phase diagram and discuss our findings in the various regimes. DMFT, which incorporates quantum fluctuations beyond mean-field theories, captures the many-body-correlation-induced NFL and Mott insulating phases and highlights the contribution of the flat band as well. Lieb theorem of the ferromagnetism holds true only for small temperatures, as evident in our findings. There is a smooth crossover from weak coupling ferromagnetic to strong coupling ferrimagnetic behavior with varying interaction. The linear behavior of the critical temperature with varying $U$ in weak coupling regime is congruous with Ref.~\cite{PhysRevB.83.220503}. The finite temperature non-magnetic NFL regime is the concomitant of the flat band singularity, as shown by the non-analytic structure of the local self-energy. The stripe order in the doped regime for the $2$D Lieb lattice is one of the key findings of this work and can be related to FFLO phase of the attractive $U$ Hubbard model. The stripe order is stable for comparatively smaller interaction strengths than on the square lattice~\cite{PhysRevB.89.155134}.

In the present work, we only consider the local self-energy approximation where spatial fluctuations are ignored. We have also carried out cellular DMFT~\cite{RevModPhys.77.1027} calculations at half-filling with a three site cluster and the results are in agreement with the R-DMFT calculation for small $U$ values, where the flat band contribution is significant. The local magnetization obtained using R-DMFT is consistent with the cellular DMFT for moderate interaction strengths while it deviates quantitatively only for large $U$ at a given $T$. This suggest that the phase diagrams obtained from both methods are qualitatively similar. The quantitative deviation can be due to non-local correlations which get significant for large interactions. We prefer the R-DMFT approach over the cellular DMFT in the doped case for two reasons: The method gets computationally extravagant due to the large unit cells needed to capture the stripe order and the QMC method has an inherent sign problem away from half-filling.

There have been a few real materials~\cite{doi:10.1143/PTP.99.489, PhysRevB.91.165137} and some density functional theory (DFT) predictions \cite{PhysRevLett.115.156401, hausoel2017local} displaying a flat band dispersion and its signature on different interaction induced instabilities, e.g. magnetism and superconductivity. Our findings about the repulsive Hubbard model on the Lieb lattice can be relevant to such materials. The high controllabity and tunability of the ultra-cold atom systems combined with the possibility of studying magnetism and Mott transitions in 2D systems~\cite{doi:10.1146/annurev-conmatphys-070909-104059}, is promising for the realization of the Hubbard model on the Lieb lattice in the near future.

\section*{Acknowledgments} 
We thank Long Liang and S. Peotta for useful discussions. This work was supported by the Academy of Finland through its Centers of Excellence Programme (2012-2017) and under Project Nos. 284621, 303351 and 307419, and by the European Research Council (ERC-2013-AdG-340748-CODE). T.I.V.
acknowledges the support from the V\"ais\"al\"a  foundation. Computing resources were provided by CSC - the Finnish IT Centre for Science.

\bibliographystyle{apsrev4-1} 
\bibliography{apssamp}

\begin{thebibliography}{76}%
\makeatletter
\providecommand \@ifxundefined [1]{%
 \@ifx{#1\undefined}
}%
\providecommand \@ifnum [1]{%
 \ifnum #1\expandafter \@firstoftwo
 \else \expandafter \@secondoftwo
 \fi
}%
\providecommand \@ifx [1]{%
 \ifx #1\expandafter \@firstoftwo
 \else \expandafter \@secondoftwo
 \fi
}%
\providecommand \natexlab [1]{#1}%
\providecommand \enquote  [1]{``#1''}%
\providecommand \bibnamefont  [1]{#1}%
\providecommand \bibfnamefont [1]{#1}%
\providecommand \citenamefont [1]{#1}%
\providecommand \href@noop [0]{\@secondoftwo}%
\providecommand \href [0]{\begingroup \@sanitize@url \@href}%
\providecommand \@href[1]{\@@startlink{#1}\@@href}%
\providecommand \@@href[1]{\endgroup#1\@@endlink}%
\providecommand \@sanitize@url [0]{\catcode `\\12\catcode `\$12\catcode
  `\&12\catcode `\#12\catcode `\^12\catcode `\_12\catcode `\%12\relax}%
\providecommand \@@startlink[1]{}%
\providecommand \@@endlink[0]{}%
\providecommand \url  [0]{\begingroup\@sanitize@url \@url }%
\providecommand \@url [1]{\endgroup\@href {#1}{\urlprefix }}%
\providecommand \urlprefix  [0]{URL }%
\providecommand \Eprint [0]{\href }%
\providecommand \doibase [0]{http://dx.doi.org/}%
\providecommand \selectlanguage [0]{\@gobble}%
\providecommand \bibinfo  [0]{\@secondoftwo}%
\providecommand \bibfield  [0]{\@secondoftwo}%
\providecommand \translation [1]{[#1]}%
\providecommand \BibitemOpen [0]{}%
\providecommand \bibitemStop [0]{}%
\providecommand \bibitemNoStop [0]{.\EOS\space}%
\providecommand \EOS [0]{\spacefactor3000\relax}%
\providecommand \BibitemShut  [1]{\csname bibitem#1\endcsname}%
\let\auto@bib@innerbib\@empty
\bibitem [{\citenamefont {Hlubina}\ \emph {et~al.}(1997)\citenamefont
  {Hlubina}, \citenamefont {Sorella},\ and\ \citenamefont
  {Guinea}}]{PhysRevLett.78.1343}%
  \BibitemOpen
  \bibfield  {author} {\bibinfo {author} {\bibfnamefont {R.}~\bibnamefont
  {Hlubina}}, \bibinfo {author} {\bibfnamefont {S.}~\bibnamefont {Sorella}}, \
  and\ \bibinfo {author} {\bibfnamefont {F.}~\bibnamefont {Guinea}},\ }\href
  {\doibase 10.1103/PhysRevLett.78.1343} {\bibfield  {journal} {\bibinfo
  {journal} {Phys. Rev. Lett.}\ }\textbf {\bibinfo {volume} {78}},\ \bibinfo
  {pages} {1343} (\bibinfo {year} {1997})}\BibitemShut {NoStop}%
\bibitem [{\citenamefont {Hirsch}\ and\ \citenamefont
  {Scalapino}(1986)}]{PhysRevLett.56.2732}%
  \BibitemOpen
  \bibfield  {author} {\bibinfo {author} {\bibfnamefont {J.~E.}\ \bibnamefont
  {Hirsch}}\ and\ \bibinfo {author} {\bibfnamefont {D.~J.}\ \bibnamefont
  {Scalapino}},\ }\href {\doibase 10.1103/PhysRevLett.56.2732} {\bibfield
  {journal} {\bibinfo  {journal} {Phys. Rev. Lett.}\ }\textbf {\bibinfo
  {volume} {56}},\ \bibinfo {pages} {2732} (\bibinfo {year}
  {1986})}\BibitemShut {NoStop}%
\bibitem [{\citenamefont {\ifmmode~\check{Z}\else \v{Z}\fi{}itko}\ \emph
  {et~al.}(2009)\citenamefont {\ifmmode~\check{Z}\else \v{Z}\fi{}itko},
  \citenamefont {Bon\ifmmode~\check{c}\else \v{c}\fi{}a},\ and\ \citenamefont
  {Pruschke}}]{PhysRevB.80.245112}%
  \BibitemOpen
  \bibfield  {author} {\bibinfo {author} {\bibfnamefont {R.}~\bibnamefont
  {\ifmmode~\check{Z}\else \v{Z}\fi{}itko}}, \bibinfo {author} {\bibfnamefont
  {J.}~\bibnamefont {Bon\ifmmode~\check{c}\else \v{c}\fi{}a}}, \ and\ \bibinfo
  {author} {\bibfnamefont {T.}~\bibnamefont {Pruschke}},\ }\href {\doibase
  10.1103/PhysRevB.80.245112} {\bibfield  {journal} {\bibinfo  {journal} {Phys.
  Rev. B}\ }\textbf {\bibinfo {volume} {80}},\ \bibinfo {pages} {245112}
  (\bibinfo {year} {2009})}\BibitemShut {NoStop}%
\bibitem [{\citenamefont {Tasaki}(1998)}]{doi:10.1143/PTP.99.489}%
  \BibitemOpen
  \bibfield  {author} {\bibinfo {author} {\bibfnamefont {H.}~\bibnamefont
  {Tasaki}},\ }\href {\doibase 10.1143/PTP.99.489} {\bibfield  {journal}
  {\bibinfo  {journal} {Progress of Theoretical Physics}\ }\textbf {\bibinfo
  {volume} {99}},\ \bibinfo {pages} {489} (\bibinfo {year} {1998})}\BibitemShut
  {NoStop}%
\bibitem [{\citenamefont {Lieb}(1989)}]{PhysRevLett.62.1201}%
  \BibitemOpen
  \bibfield  {author} {\bibinfo {author} {\bibfnamefont {E.~H.}\ \bibnamefont
  {Lieb}},\ }\href {\doibase 10.1103/PhysRevLett.62.1201} {\bibfield  {journal}
  {\bibinfo  {journal} {Phys. Rev. Lett.}\ }\textbf {\bibinfo {volume} {62}},\
  \bibinfo {pages} {1201} (\bibinfo {year} {1989})}\BibitemShut {NoStop}%
\bibitem [{\citenamefont {Noda}\ \emph {et~al.}(2009)\citenamefont {Noda},
  \citenamefont {Koga}, \citenamefont {Kawakami},\ and\ \citenamefont
  {Pruschke}}]{PhysRevA.80.063622}%
  \BibitemOpen
  \bibfield  {author} {\bibinfo {author} {\bibfnamefont {K.}~\bibnamefont
  {Noda}}, \bibinfo {author} {\bibfnamefont {A.}~\bibnamefont {Koga}}, \bibinfo
  {author} {\bibfnamefont {N.}~\bibnamefont {Kawakami}}, \ and\ \bibinfo
  {author} {\bibfnamefont {T.}~\bibnamefont {Pruschke}},\ }\href {\doibase
  10.1103/PhysRevA.80.063622} {\bibfield  {journal} {\bibinfo  {journal} {Phys.
  Rev. A}\ }\textbf {\bibinfo {volume} {80}},\ \bibinfo {pages} {063622}
  (\bibinfo {year} {2009})}\BibitemShut {NoStop}%
\bibitem [{\citenamefont {Hartman}\ \emph {et~al.}(2016)\citenamefont
  {Hartman}, \citenamefont {Chiu},\ and\ \citenamefont
  {Scalettar}}]{PhysRevB.93.235143}%
  \BibitemOpen
  \bibfield  {author} {\bibinfo {author} {\bibfnamefont {N.}~\bibnamefont
  {Hartman}}, \bibinfo {author} {\bibfnamefont {W.-T.}\ \bibnamefont {Chiu}}, \
  and\ \bibinfo {author} {\bibfnamefont {R.~T.}\ \bibnamefont {Scalettar}},\
  }\href {\doibase 10.1103/PhysRevB.93.235143} {\bibfield  {journal} {\bibinfo
  {journal} {Phys. Rev. B}\ }\textbf {\bibinfo {volume} {93}},\ \bibinfo
  {pages} {235143} (\bibinfo {year} {2016})}\BibitemShut {NoStop}%
\bibitem [{\citenamefont {Arita}\ \emph {et~al.}(2002)\citenamefont {Arita},
  \citenamefont {Suwa}, \citenamefont {Kuroki},\ and\ \citenamefont
  {Aoki}}]{PhysRevLett.88.127202}%
  \BibitemOpen
  \bibfield  {author} {\bibinfo {author} {\bibfnamefont {R.}~\bibnamefont
  {Arita}}, \bibinfo {author} {\bibfnamefont {Y.}~\bibnamefont {Suwa}},
  \bibinfo {author} {\bibfnamefont {K.}~\bibnamefont {Kuroki}}, \ and\ \bibinfo
  {author} {\bibfnamefont {H.}~\bibnamefont {Aoki}},\ }\href {\doibase
  10.1103/PhysRevLett.88.127202} {\bibfield  {journal} {\bibinfo  {journal}
  {Phys. Rev. Lett.}\ }\textbf {\bibinfo {volume} {88}},\ \bibinfo {pages}
  {127202} (\bibinfo {year} {2002})}\BibitemShut {NoStop}%
\bibitem [{\citenamefont {Julku}\ \emph {et~al.}(2016)\citenamefont {Julku},
  \citenamefont {Peotta}, \citenamefont {Vanhala}, \citenamefont {Kim},\ and\
  \citenamefont {T\"orm\"a}}]{PhysRevLett.117.045303}%
  \BibitemOpen
  \bibfield  {author} {\bibinfo {author} {\bibfnamefont {A.}~\bibnamefont
  {Julku}}, \bibinfo {author} {\bibfnamefont {S.}~\bibnamefont {Peotta}},
  \bibinfo {author} {\bibfnamefont {T.~I.}\ \bibnamefont {Vanhala}}, \bibinfo
  {author} {\bibfnamefont {D.-H.}\ \bibnamefont {Kim}}, \ and\ \bibinfo
  {author} {\bibfnamefont {P.}~\bibnamefont {T\"orm\"a}},\ }\href {\doibase
  10.1103/PhysRevLett.117.045303} {\bibfield  {journal} {\bibinfo  {journal}
  {Phys. Rev. Lett.}\ }\textbf {\bibinfo {volume} {117}},\ \bibinfo {pages}
  {045303} (\bibinfo {year} {2016})}\BibitemShut {NoStop}%
\bibitem [{\citenamefont {Peotta}\ and\ \citenamefont
  {T{\"o}rm{\"a}}(2015)}]{Peotta2015}%
  \BibitemOpen
  \bibfield  {author} {\bibinfo {author} {\bibfnamefont {S.}~\bibnamefont
  {Peotta}}\ and\ \bibinfo {author} {\bibfnamefont {P.}~\bibnamefont
  {T{\"o}rm{\"a}}},\ }\href@noop {} {\bibfield  {journal} {\bibinfo  {journal}
  {Nature Communications}\ }\textbf {\bibinfo {volume} {6}},\ \bibinfo {pages}
  {8944} (\bibinfo {year} {2015})}\BibitemShut {NoStop}%
\bibitem [{\citenamefont {Liang}\ \emph {et~al.}(2017)\citenamefont {Liang},
  \citenamefont {Vanhala}, \citenamefont {Peotta}, \citenamefont {Siro},
  \citenamefont {Harju},\ and\ \citenamefont {T\"orm\"a}}]{PhysRevB.95.024515}%
  \BibitemOpen
  \bibfield  {author} {\bibinfo {author} {\bibfnamefont {L.}~\bibnamefont
  {Liang}}, \bibinfo {author} {\bibfnamefont {T.~I.}\ \bibnamefont {Vanhala}},
  \bibinfo {author} {\bibfnamefont {S.}~\bibnamefont {Peotta}}, \bibinfo
  {author} {\bibfnamefont {T.}~\bibnamefont {Siro}}, \bibinfo {author}
  {\bibfnamefont {A.}~\bibnamefont {Harju}}, \ and\ \bibinfo {author}
  {\bibfnamefont {P.}~\bibnamefont {T\"orm\"a}},\ }\href {\doibase
  10.1103/PhysRevB.95.024515} {\bibfield  {journal} {\bibinfo  {journal} {Phys.
  Rev. B}\ }\textbf {\bibinfo {volume} {95}},\ \bibinfo {pages} {024515}
  (\bibinfo {year} {2017})}\BibitemShut {NoStop}%
\bibitem [{\citenamefont {Khodel}\ \emph {et~al.}(2015)\citenamefont {Khodel},
  \citenamefont {Clark}, \citenamefont {Popov},\ and\ \citenamefont
  {Shaginyan}}]{Khodel2015}%
  \BibitemOpen
  \bibfield  {author} {\bibinfo {author} {\bibfnamefont {V.~A.}\ \bibnamefont
  {Khodel}}, \bibinfo {author} {\bibfnamefont {J.~W.}\ \bibnamefont {Clark}},
  \bibinfo {author} {\bibfnamefont {K.~G.}\ \bibnamefont {Popov}}, \ and\
  \bibinfo {author} {\bibfnamefont {V.~R.}\ \bibnamefont {Shaginyan}},\ }\href
  {\doibase 10.1134/S0021364015060065} {\bibfield  {journal} {\bibinfo
  {journal} {JETP Letters}\ }\textbf {\bibinfo {volume} {101}},\ \bibinfo
  {pages} {413} (\bibinfo {year} {2015})}\BibitemShut {NoStop}%
\bibitem [{\citenamefont {Shinaoka}\ \emph {et~al.}(2015)\citenamefont
  {Shinaoka}, \citenamefont {Hoshino}, \citenamefont {Troyer},\ and\
  \citenamefont {Werner}}]{PhysRevLett.115.156401}%
  \BibitemOpen
  \bibfield  {author} {\bibinfo {author} {\bibfnamefont {H.}~\bibnamefont
  {Shinaoka}}, \bibinfo {author} {\bibfnamefont {S.}~\bibnamefont {Hoshino}},
  \bibinfo {author} {\bibfnamefont {M.}~\bibnamefont {Troyer}}, \ and\ \bibinfo
  {author} {\bibfnamefont {P.}~\bibnamefont {Werner}},\ }\href {\doibase
  10.1103/PhysRevLett.115.156401} {\bibfield  {journal} {\bibinfo  {journal}
  {Phys. Rev. Lett.}\ }\textbf {\bibinfo {volume} {115}},\ \bibinfo {pages}
  {156401} (\bibinfo {year} {2015})}\BibitemShut {NoStop}%
\bibitem [{\citenamefont {Hausoel}\ \emph {et~al.}(2017)\citenamefont
  {Hausoel}, \citenamefont {Karolak}, \citenamefont {Sasɩoglu}, \citenamefont
  {Lichtenstein}, \citenamefont {Held}, \citenamefont {Katanin}, \citenamefont
  {Toschi},\ and\ \citenamefont {Sangiovanni}}]{hausoel2017local}%
  \BibitemOpen
  \bibfield  {author} {\bibinfo {author} {\bibfnamefont {A.}~\bibnamefont
  {Hausoel}}, \bibinfo {author} {\bibfnamefont {M.}~\bibnamefont {Karolak}},
  \bibinfo {author} {\bibfnamefont {E.}~\bibnamefont {Sasɩoglu}}, \bibinfo
  {author} {\bibfnamefont {A.}~\bibnamefont {Lichtenstein}}, \bibinfo {author}
  {\bibfnamefont {K.}~\bibnamefont {Held}}, \bibinfo {author} {\bibfnamefont
  {A.}~\bibnamefont {Katanin}}, \bibinfo {author} {\bibfnamefont
  {A.}~\bibnamefont {Toschi}}, \ and\ \bibinfo {author} {\bibfnamefont
  {G.}~\bibnamefont {Sangiovanni}},\ }\href@noop {} {\bibfield  {journal}
  {\bibinfo  {journal} {Nature Communications}\ }\textbf {\bibinfo {volume}
  {8}} (\bibinfo {year} {2017})}\BibitemShut {NoStop}%
\bibitem [{\citenamefont {Goldman}\ \emph {et~al.}(2011)\citenamefont
  {Goldman}, \citenamefont {Urban},\ and\ \citenamefont
  {Bercioux}}]{PhysRevA.83.063601}%
  \BibitemOpen
  \bibfield  {author} {\bibinfo {author} {\bibfnamefont {N.}~\bibnamefont
  {Goldman}}, \bibinfo {author} {\bibfnamefont {D.~F.}\ \bibnamefont {Urban}},
  \ and\ \bibinfo {author} {\bibfnamefont {D.}~\bibnamefont {Bercioux}},\
  }\href {\doibase 10.1103/PhysRevA.83.063601} {\bibfield  {journal} {\bibinfo
  {journal} {Phys. Rev. A}\ }\textbf {\bibinfo {volume} {83}},\ \bibinfo
  {pages} {063601} (\bibinfo {year} {2011})}\BibitemShut {NoStop}%
\bibitem [{\citenamefont {Ochi}\ \emph {et~al.}(2015)\citenamefont {Ochi},
  \citenamefont {Arita}, \citenamefont {Matsumoto}, \citenamefont {Kino},\ and\
  \citenamefont {Miyake}}]{PhysRevB.91.165137}%
  \BibitemOpen
  \bibfield  {author} {\bibinfo {author} {\bibfnamefont {M.}~\bibnamefont
  {Ochi}}, \bibinfo {author} {\bibfnamefont {R.}~\bibnamefont {Arita}},
  \bibinfo {author} {\bibfnamefont {M.}~\bibnamefont {Matsumoto}}, \bibinfo
  {author} {\bibfnamefont {H.}~\bibnamefont {Kino}}, \ and\ \bibinfo {author}
  {\bibfnamefont {T.}~\bibnamefont {Miyake}},\ }\href {\doibase
  10.1103/PhysRevB.91.165137} {\bibfield  {journal} {\bibinfo  {journal} {Phys.
  Rev. B}\ }\textbf {\bibinfo {volume} {91}},\ \bibinfo {pages} {165137}
  (\bibinfo {year} {2015})}\BibitemShut {NoStop}%
\bibitem [{\citenamefont {Taie}\ \emph {et~al.}(2015)\citenamefont {Taie},
  \citenamefont {Ozawa}, \citenamefont {Ichinose}, \citenamefont {Nishio},
  \citenamefont {Nakajima},\ and\ \citenamefont {Takahashi}}]{Taiee1500854}%
  \BibitemOpen
  \bibfield  {author} {\bibinfo {author} {\bibfnamefont {S.}~\bibnamefont
  {Taie}}, \bibinfo {author} {\bibfnamefont {H.}~\bibnamefont {Ozawa}},
  \bibinfo {author} {\bibfnamefont {T.}~\bibnamefont {Ichinose}}, \bibinfo
  {author} {\bibfnamefont {T.}~\bibnamefont {Nishio}}, \bibinfo {author}
  {\bibfnamefont {S.}~\bibnamefont {Nakajima}}, \ and\ \bibinfo {author}
  {\bibfnamefont {Y.}~\bibnamefont {Takahashi}},\ }\href {\doibase
  10.1126/sciadv.1500854} {\bibfield  {journal} {\bibinfo  {journal} {Science
  Advances}\ }\textbf {\bibinfo {volume} {1}} (\bibinfo {year} {2015}),\
  10.1126/sciadv.1500854}\BibitemShut {NoStop}%
\bibitem [{\citenamefont {Greif}\ \emph {et~al.}(2016)\citenamefont {Greif},
  \citenamefont {Parsons}, \citenamefont {Mazurenko}, \citenamefont {Chiu},
  \citenamefont {Blatt}, \citenamefont {Huber}, \citenamefont {Ji},\ and\
  \citenamefont {Greiner}}]{Greif953}%
  \BibitemOpen
  \bibfield  {author} {\bibinfo {author} {\bibfnamefont {D.}~\bibnamefont
  {Greif}}, \bibinfo {author} {\bibfnamefont {M.~F.}\ \bibnamefont {Parsons}},
  \bibinfo {author} {\bibfnamefont {A.}~\bibnamefont {Mazurenko}}, \bibinfo
  {author} {\bibfnamefont {C.~S.}\ \bibnamefont {Chiu}}, \bibinfo {author}
  {\bibfnamefont {S.}~\bibnamefont {Blatt}}, \bibinfo {author} {\bibfnamefont
  {F.}~\bibnamefont {Huber}}, \bibinfo {author} {\bibfnamefont
  {G.}~\bibnamefont {Ji}}, \ and\ \bibinfo {author} {\bibfnamefont
  {M.}~\bibnamefont {Greiner}},\ }\href {\doibase 10.1126/science.aad9041}
  {\bibfield  {journal} {\bibinfo  {journal} {Science}\ }\textbf {\bibinfo
  {volume} {351}},\ \bibinfo {pages} {953} (\bibinfo {year}
  {2016})}\BibitemShut {NoStop}%
\bibitem [{\citenamefont {Bloch}\ \emph {et~al.}(2008)\citenamefont {Bloch},
  \citenamefont {Dalibard},\ and\ \citenamefont {Zwerger}}]{RevModPhys.80.885}%
  \BibitemOpen
  \bibfield  {author} {\bibinfo {author} {\bibfnamefont {I.}~\bibnamefont
  {Bloch}}, \bibinfo {author} {\bibfnamefont {J.}~\bibnamefont {Dalibard}}, \
  and\ \bibinfo {author} {\bibfnamefont {W.}~\bibnamefont {Zwerger}},\ }\href
  {\doibase 10.1103/RevModPhys.80.885} {\bibfield  {journal} {\bibinfo
  {journal} {Rev. Mod. Phys.}\ }\textbf {\bibinfo {volume} {80}},\ \bibinfo
  {pages} {885} (\bibinfo {year} {2008})}\BibitemShut {NoStop}%
\bibitem [{\citenamefont
  {Esslinger}(2010)}]{doi:10.1146/annurev-conmatphys-070909-104059}%
  \BibitemOpen
  \bibfield  {author} {\bibinfo {author} {\bibfnamefont {T.}~\bibnamefont
  {Esslinger}},\ }\href {\doibase 10.1146/annurev-conmatphys-070909-104059}
  {\bibfield  {journal} {\bibinfo  {journal} {Annual Review of Condensed Matter
  Physics}\ }\textbf {\bibinfo {volume} {1}},\ \bibinfo {pages} {129} (\bibinfo
  {year} {2010})}\BibitemShut {NoStop}%
\bibitem [{\citenamefont {Miranda}\ and\ \citenamefont
  {Dobrosavljević}(2005)}]{0034-4885-68-10-R02}%
  \BibitemOpen
  \bibfield  {author} {\bibinfo {author} {\bibfnamefont {E.}~\bibnamefont
  {Miranda}}\ and\ \bibinfo {author} {\bibfnamefont {V.}~\bibnamefont
  {Dobrosavljević}},\ }\href {http://stacks.iop.org/0034-4885/68/i=10/a=R02}
  {\bibfield  {journal} {\bibinfo  {journal} {Reports on Progress in Physics}\
  }\textbf {\bibinfo {volume} {68}},\ \bibinfo {pages} {2337} (\bibinfo {year}
  {2005})}\BibitemShut {NoStop}%
\bibitem [{\citenamefont {Vidhyadhiraja}\ and\ \citenamefont
  {Kumar}(2013)}]{PhysRevB.88.195120}%
  \BibitemOpen
  \bibfield  {author} {\bibinfo {author} {\bibfnamefont {N.~S.}\ \bibnamefont
  {Vidhyadhiraja}}\ and\ \bibinfo {author} {\bibfnamefont {P.}~\bibnamefont
  {Kumar}},\ }\href {\doibase 10.1103/PhysRevB.88.195120} {\bibfield  {journal}
  {\bibinfo  {journal} {Phys. Rev. B}\ }\textbf {\bibinfo {volume} {88}},\
  \bibinfo {pages} {195120} (\bibinfo {year} {2013})}\BibitemShut {NoStop}%
\bibitem [{\citenamefont {Vojta}(2012)}]{VOJTA2012178}%
  \BibitemOpen
  \bibfield  {author} {\bibinfo {author} {\bibfnamefont {M.}~\bibnamefont
  {Vojta}},\ }\href {\doibase http://dx.doi.org/10.1016/j.physc.2012.04.013}
  {\bibfield  {journal} {\bibinfo  {journal} {Physica C: Superconductivity}\
  }\textbf {\bibinfo {volume} {481}},\ \bibinfo {pages} {178 } (\bibinfo {year}
  {2012})}\BibitemShut {NoStop}%
\bibitem [{\citenamefont {Sun}\ \emph {et~al.}(2011)\citenamefont {Sun},
  \citenamefont {Gu}, \citenamefont {Katsura},\ and\ \citenamefont
  {Das~Sarma}}]{PhysRevLett.106.236803}%
  \BibitemOpen
  \bibfield  {author} {\bibinfo {author} {\bibfnamefont {K.}~\bibnamefont
  {Sun}}, \bibinfo {author} {\bibfnamefont {Z.}~\bibnamefont {Gu}}, \bibinfo
  {author} {\bibfnamefont {H.}~\bibnamefont {Katsura}}, \ and\ \bibinfo
  {author} {\bibfnamefont {S.}~\bibnamefont {Das~Sarma}},\ }\href {\doibase
  10.1103/PhysRevLett.106.236803} {\bibfield  {journal} {\bibinfo  {journal}
  {Phys. Rev. Lett.}\ }\textbf {\bibinfo {volume} {106}},\ \bibinfo {pages}
  {236803} (\bibinfo {year} {2011})}\BibitemShut {NoStop}%
\bibitem [{\citenamefont {Takahashi}\ and\ \citenamefont
  {Murakami}(2013)}]{PhysRevB.88.235303}%
  \BibitemOpen
  \bibfield  {author} {\bibinfo {author} {\bibfnamefont {R.}~\bibnamefont
  {Takahashi}}\ and\ \bibinfo {author} {\bibfnamefont {S.}~\bibnamefont
  {Murakami}},\ }\href {\doibase 10.1103/PhysRevB.88.235303} {\bibfield
  {journal} {\bibinfo  {journal} {Phys. Rev. B}\ }\textbf {\bibinfo {volume}
  {88}},\ \bibinfo {pages} {235303} (\bibinfo {year} {2013})}\BibitemShut
  {NoStop}%
\bibitem [{\citenamefont {Ozawa}\ \emph {et~al.}(2017)\citenamefont {Ozawa},
  \citenamefont {Taie}, \citenamefont {Ichinose},\ and\ \citenamefont
  {Takahashi}}]{PhysRevLett.118.175301}%
  \BibitemOpen
  \bibfield  {author} {\bibinfo {author} {\bibfnamefont {H.}~\bibnamefont
  {Ozawa}}, \bibinfo {author} {\bibfnamefont {S.}~\bibnamefont {Taie}},
  \bibinfo {author} {\bibfnamefont {T.}~\bibnamefont {Ichinose}}, \ and\
  \bibinfo {author} {\bibfnamefont {Y.}~\bibnamefont {Takahashi}},\ }\href
  {\doibase 10.1103/PhysRevLett.118.175301} {\bibfield  {journal} {\bibinfo
  {journal} {Phys. Rev. Lett.}\ }\textbf {\bibinfo {volume} {118}},\ \bibinfo
  {pages} {175301} (\bibinfo {year} {2017})}\BibitemShut {NoStop}%
\bibitem [{\citenamefont {Vicencio}\ \emph {et~al.}(2015)\citenamefont
  {Vicencio}, \citenamefont {Cantillano}, \citenamefont {Morales-Inostroza},
  \citenamefont {Real}, \citenamefont {Mej\'{\i}a-Cort\'es}, \citenamefont
  {Weimann}, \citenamefont {Szameit},\ and\ \citenamefont
  {Molina}}]{PhysRevLett.114.245503}%
  \BibitemOpen
  \bibfield  {author} {\bibinfo {author} {\bibfnamefont {R.~A.}\ \bibnamefont
  {Vicencio}}, \bibinfo {author} {\bibfnamefont {C.}~\bibnamefont
  {Cantillano}}, \bibinfo {author} {\bibfnamefont {L.}~\bibnamefont
  {Morales-Inostroza}}, \bibinfo {author} {\bibfnamefont {B.}~\bibnamefont
  {Real}}, \bibinfo {author} {\bibfnamefont {C.}~\bibnamefont
  {Mej\'{\i}a-Cort\'es}}, \bibinfo {author} {\bibfnamefont {S.}~\bibnamefont
  {Weimann}}, \bibinfo {author} {\bibfnamefont {A.}~\bibnamefont {Szameit}}, \
  and\ \bibinfo {author} {\bibfnamefont {M.~I.}\ \bibnamefont {Molina}},\
  }\href {\doibase 10.1103/PhysRevLett.114.245503} {\bibfield  {journal}
  {\bibinfo  {journal} {Phys. Rev. Lett.}\ }\textbf {\bibinfo {volume} {114}},\
  \bibinfo {pages} {245503} (\bibinfo {year} {2015})}\BibitemShut {NoStop}%
\bibitem [{\citenamefont {Mukherjee}\ \emph {et~al.}(2015)\citenamefont
  {Mukherjee}, \citenamefont {Spracklen}, \citenamefont {Choudhury},
  \citenamefont {Goldman}, \citenamefont {\"Ohberg}, \citenamefont
  {Andersson},\ and\ \citenamefont {Thomson}}]{PhysRevLett.114.245504}%
  \BibitemOpen
  \bibfield  {author} {\bibinfo {author} {\bibfnamefont {S.}~\bibnamefont
  {Mukherjee}}, \bibinfo {author} {\bibfnamefont {A.}~\bibnamefont
  {Spracklen}}, \bibinfo {author} {\bibfnamefont {D.}~\bibnamefont
  {Choudhury}}, \bibinfo {author} {\bibfnamefont {N.}~\bibnamefont {Goldman}},
  \bibinfo {author} {\bibfnamefont {P.}~\bibnamefont {\"Ohberg}}, \bibinfo
  {author} {\bibfnamefont {E.}~\bibnamefont {Andersson}}, \ and\ \bibinfo
  {author} {\bibfnamefont {R.~R.}\ \bibnamefont {Thomson}},\ }\href {\doibase
  10.1103/PhysRevLett.114.245504} {\bibfield  {journal} {\bibinfo  {journal}
  {Phys. Rev. Lett.}\ }\textbf {\bibinfo {volume} {114}},\ \bibinfo {pages}
  {245504} (\bibinfo {year} {2015})}\BibitemShut {NoStop}%
\bibitem [{\citenamefont {Slot}\ \emph {et~al.}(2017)\citenamefont {Slot},
  \citenamefont {Gardenier}, \citenamefont {Jacobse}, \citenamefont {van
  Miert}, \citenamefont {Kempkes}, \citenamefont {Zevenhuizen}, \citenamefont
  {Smith}, \citenamefont {Vanmaekelbergh},\ and\ \citenamefont
  {Swart}}]{Slot2017}%
  \BibitemOpen
  \bibfield  {author} {\bibinfo {author} {\bibfnamefont {M.~R.}\ \bibnamefont
  {Slot}}, \bibinfo {author} {\bibfnamefont {T.~S.}\ \bibnamefont {Gardenier}},
  \bibinfo {author} {\bibfnamefont {P.~H.}\ \bibnamefont {Jacobse}}, \bibinfo
  {author} {\bibfnamefont {G.~C.~P.}\ \bibnamefont {van Miert}}, \bibinfo
  {author} {\bibfnamefont {S.~N.}\ \bibnamefont {Kempkes}}, \bibinfo {author}
  {\bibfnamefont {S.~J.~M.}\ \bibnamefont {Zevenhuizen}}, \bibinfo {author}
  {\bibfnamefont {C.~M.}\ \bibnamefont {Smith}}, \bibinfo {author}
  {\bibfnamefont {D.}~\bibnamefont {Vanmaekelbergh}}, \ and\ \bibinfo {author}
  {\bibfnamefont {I.}~\bibnamefont {Swart}},\ }\href
  {http://dx.doi.org/10.1038/nphys4105} {\bibfield  {journal} {\bibinfo
  {journal} {Nat Phys}\ }\textbf {\bibinfo {volume} {13}},\ \bibinfo {pages}
  {672} (\bibinfo {year} {2017})}\BibitemShut {NoStop}%
\bibitem [{\citenamefont {Drost}\ \emph {et~al.}(2017)\citenamefont {Drost},
  \citenamefont {Ojanen}, \citenamefont {Harju},\ and\ \citenamefont
  {Liljeroth}}]{Drost2017}%
  \BibitemOpen
  \bibfield  {author} {\bibinfo {author} {\bibfnamefont {R.}~\bibnamefont
  {Drost}}, \bibinfo {author} {\bibfnamefont {T.}~\bibnamefont {Ojanen}},
  \bibinfo {author} {\bibfnamefont {A.}~\bibnamefont {Harju}}, \ and\ \bibinfo
  {author} {\bibfnamefont {P.}~\bibnamefont {Liljeroth}},\ }\href
  {http://dx.doi.org/10.1038/nphys4080} {\bibfield  {journal} {\bibinfo
  {journal} {Nat Phys}\ }\textbf {\bibinfo {volume} {13}},\ \bibinfo {pages}
  {668} (\bibinfo {year} {2017})}\BibitemShut {NoStop}%
\bibitem [{\citenamefont {Noda}\ \emph {et~al.}(2015)\citenamefont {Noda},
  \citenamefont {Inaba},\ and\ \citenamefont {Yamashita}}]{PhysRevA.91.063610}%
  \BibitemOpen
  \bibfield  {author} {\bibinfo {author} {\bibfnamefont {K.}~\bibnamefont
  {Noda}}, \bibinfo {author} {\bibfnamefont {K.}~\bibnamefont {Inaba}}, \ and\
  \bibinfo {author} {\bibfnamefont {M.}~\bibnamefont {Yamashita}},\ }\href
  {\doibase 10.1103/PhysRevA.91.063610} {\bibfield  {journal} {\bibinfo
  {journal} {Phys. Rev. A}\ }\textbf {\bibinfo {volume} {91}},\ \bibinfo
  {pages} {063610} (\bibinfo {year} {2015})}\BibitemShut {NoStop}%
\bibitem [{\citenamefont {Mermin}\ and\ \citenamefont
  {Wagner}(1966)}]{PhysRevLett.17.1133}%
  \BibitemOpen
  \bibfield  {author} {\bibinfo {author} {\bibfnamefont {N.~D.}\ \bibnamefont
  {Mermin}}\ and\ \bibinfo {author} {\bibfnamefont {H.}~\bibnamefont
  {Wagner}},\ }\href {\doibase 10.1103/PhysRevLett.17.1133} {\bibfield
  {journal} {\bibinfo  {journal} {Phys. Rev. Lett.}\ }\textbf {\bibinfo
  {volume} {17}},\ \bibinfo {pages} {1133} (\bibinfo {year}
  {1966})}\BibitemShut {NoStop}%
\bibitem [{\citenamefont {Hohenberg}(1967)}]{PhysRev.158.383}%
  \BibitemOpen
  \bibfield  {author} {\bibinfo {author} {\bibfnamefont {P.~C.}\ \bibnamefont
  {Hohenberg}},\ }\href {\doibase 10.1103/PhysRev.158.383} {\bibfield
  {journal} {\bibinfo  {journal} {Phys. Rev.}\ }\textbf {\bibinfo {volume}
  {158}},\ \bibinfo {pages} {383} (\bibinfo {year} {1967})}\BibitemShut
  {NoStop}%
\bibitem [{\citenamefont {Paiva}\ \emph {et~al.}(2010)\citenamefont {Paiva},
  \citenamefont {Scalettar}, \citenamefont {Randeria},\ and\ \citenamefont
  {Trivedi}}]{PhysRevLett.104.066406}%
  \BibitemOpen
  \bibfield  {author} {\bibinfo {author} {\bibfnamefont {T.}~\bibnamefont
  {Paiva}}, \bibinfo {author} {\bibfnamefont {R.}~\bibnamefont {Scalettar}},
  \bibinfo {author} {\bibfnamefont {M.}~\bibnamefont {Randeria}}, \ and\
  \bibinfo {author} {\bibfnamefont {N.}~\bibnamefont {Trivedi}},\ }\href
  {\doibase 10.1103/PhysRevLett.104.066406} {\bibfield  {journal} {\bibinfo
  {journal} {Phys. Rev. Lett.}\ }\textbf {\bibinfo {volume} {104}},\ \bibinfo
  {pages} {066406} (\bibinfo {year} {2010})}\BibitemShut {NoStop}%
\bibitem [{\citenamefont {Cheuk}\ \emph {et~al.}(2016)\citenamefont {Cheuk},
  \citenamefont {Nichols}, \citenamefont {Lawrence}, \citenamefont {Okan},
  \citenamefont {Zhang}, \citenamefont {Khatami}, \citenamefont {Trivedi},
  \citenamefont {Paiva}, \citenamefont {Rigol},\ and\ \citenamefont
  {Zwierlein}}]{Cheuk1260}%
  \BibitemOpen
  \bibfield  {author} {\bibinfo {author} {\bibfnamefont {L.~W.}\ \bibnamefont
  {Cheuk}}, \bibinfo {author} {\bibfnamefont {M.~A.}\ \bibnamefont {Nichols}},
  \bibinfo {author} {\bibfnamefont {K.~R.}\ \bibnamefont {Lawrence}}, \bibinfo
  {author} {\bibfnamefont {M.}~\bibnamefont {Okan}}, \bibinfo {author}
  {\bibfnamefont {H.}~\bibnamefont {Zhang}}, \bibinfo {author} {\bibfnamefont
  {E.}~\bibnamefont {Khatami}}, \bibinfo {author} {\bibfnamefont
  {N.}~\bibnamefont {Trivedi}}, \bibinfo {author} {\bibfnamefont
  {T.}~\bibnamefont {Paiva}}, \bibinfo {author} {\bibfnamefont
  {M.}~\bibnamefont {Rigol}}, \ and\ \bibinfo {author} {\bibfnamefont {M.~W.}\
  \bibnamefont {Zwierlein}},\ }\href {\doibase 10.1126/science.aag3349}
  {\bibfield  {journal} {\bibinfo  {journal} {Science}\ }\textbf {\bibinfo
  {volume} {353}},\ \bibinfo {pages} {1260} (\bibinfo {year}
  {2016})}\BibitemShut {NoStop}%
\bibitem [{\citenamefont {Parsons}\ \emph {et~al.}(2016)\citenamefont
  {Parsons}, \citenamefont {Mazurenko}, \citenamefont {Chiu}, \citenamefont
  {Ji}, \citenamefont {Greif},\ and\ \citenamefont {Greiner}}]{Parsons1253}%
  \BibitemOpen
  \bibfield  {author} {\bibinfo {author} {\bibfnamefont {M.~F.}\ \bibnamefont
  {Parsons}}, \bibinfo {author} {\bibfnamefont {A.}~\bibnamefont {Mazurenko}},
  \bibinfo {author} {\bibfnamefont {C.~S.}\ \bibnamefont {Chiu}}, \bibinfo
  {author} {\bibfnamefont {G.}~\bibnamefont {Ji}}, \bibinfo {author}
  {\bibfnamefont {D.}~\bibnamefont {Greif}}, \ and\ \bibinfo {author}
  {\bibfnamefont {M.}~\bibnamefont {Greiner}},\ }\href {\doibase
  10.1126/science.aag1430} {\bibfield  {journal} {\bibinfo  {journal}
  {Science}\ }\textbf {\bibinfo {volume} {353}},\ \bibinfo {pages} {1253}
  (\bibinfo {year} {2016})}\BibitemShut {NoStop}%
\bibitem [{\citenamefont {Mazurenko}\ \emph {et~al.}(2017)\citenamefont
  {Mazurenko}, \citenamefont {Chiu}, \citenamefont {Ji}, \citenamefont
  {Parsons}, \citenamefont {Kan{\'a}sz-Nagy}, \citenamefont {Schmidt},
  \citenamefont {Grusdt}, \citenamefont {Demler}, \citenamefont {Greif},\ and\
  \citenamefont {Greiner}}]{mazurenko2017cold}%
  \BibitemOpen
  \bibfield  {author} {\bibinfo {author} {\bibfnamefont {A.}~\bibnamefont
  {Mazurenko}}, \bibinfo {author} {\bibfnamefont {C.~S.}\ \bibnamefont {Chiu}},
  \bibinfo {author} {\bibfnamefont {G.}~\bibnamefont {Ji}}, \bibinfo {author}
  {\bibfnamefont {M.~F.}\ \bibnamefont {Parsons}}, \bibinfo {author}
  {\bibfnamefont {M.}~\bibnamefont {Kan{\'a}sz-Nagy}}, \bibinfo {author}
  {\bibfnamefont {R.}~\bibnamefont {Schmidt}}, \bibinfo {author} {\bibfnamefont
  {F.}~\bibnamefont {Grusdt}}, \bibinfo {author} {\bibfnamefont
  {E.}~\bibnamefont {Demler}}, \bibinfo {author} {\bibfnamefont
  {D.}~\bibnamefont {Greif}}, \ and\ \bibinfo {author} {\bibfnamefont
  {M.}~\bibnamefont {Greiner}},\ }\href {\doibase doi:10.1038/nature22362}
  {\bibfield  {journal} {\bibinfo  {journal} {Nature}\ }\textbf {\bibinfo
  {volume} {545}},\ \bibinfo {pages} {462} (\bibinfo {year}
  {2017})}\BibitemShut {NoStop}%
\bibitem [{\citenamefont {Greif}\ \emph {et~al.}(2013)\citenamefont {Greif},
  \citenamefont {Uehlinger}, \citenamefont {Jotzu}, \citenamefont {Tarruell},\
  and\ \citenamefont {Esslinger}}]{Greif1236362}%
  \BibitemOpen
  \bibfield  {author} {\bibinfo {author} {\bibfnamefont {D.}~\bibnamefont
  {Greif}}, \bibinfo {author} {\bibfnamefont {T.}~\bibnamefont {Uehlinger}},
  \bibinfo {author} {\bibfnamefont {G.}~\bibnamefont {Jotzu}}, \bibinfo
  {author} {\bibfnamefont {L.}~\bibnamefont {Tarruell}}, \ and\ \bibinfo
  {author} {\bibfnamefont {T.}~\bibnamefont {Esslinger}},\ }\href {\doibase
  10.1126/science.1236362} {\bibfield  {journal} {\bibinfo  {journal}
  {Science}\ } (\bibinfo {year} {2013}),\ 10.1126/science.1236362}\BibitemShut
  {NoStop}%
\bibitem [{\citenamefont {Drewes}\ \emph {et~al.}(2017)\citenamefont {Drewes},
  \citenamefont {Miller}, \citenamefont {Cocchi}, \citenamefont {Chan},
  \citenamefont {Wurz}, \citenamefont {Gall}, \citenamefont {Pertot},
  \citenamefont {Brennecke},\ and\ \citenamefont
  {K\"ohl}}]{PhysRevLett.118.170401}%
  \BibitemOpen
  \bibfield  {author} {\bibinfo {author} {\bibfnamefont {J.~H.}\ \bibnamefont
  {Drewes}}, \bibinfo {author} {\bibfnamefont {L.~A.}\ \bibnamefont {Miller}},
  \bibinfo {author} {\bibfnamefont {E.}~\bibnamefont {Cocchi}}, \bibinfo
  {author} {\bibfnamefont {C.~F.}\ \bibnamefont {Chan}}, \bibinfo {author}
  {\bibfnamefont {N.}~\bibnamefont {Wurz}}, \bibinfo {author} {\bibfnamefont
  {M.}~\bibnamefont {Gall}}, \bibinfo {author} {\bibfnamefont {D.}~\bibnamefont
  {Pertot}}, \bibinfo {author} {\bibfnamefont {F.}~\bibnamefont {Brennecke}}, \
  and\ \bibinfo {author} {\bibfnamefont {M.}~\bibnamefont {K\"ohl}},\ }\href
  {\doibase 10.1103/PhysRevLett.118.170401} {\bibfield  {journal} {\bibinfo
  {journal} {Phys. Rev. Lett.}\ }\textbf {\bibinfo {volume} {118}},\ \bibinfo
  {pages} {170401} (\bibinfo {year} {2017})}\BibitemShut {NoStop}%
\bibitem [{\citenamefont {Brown}\ \emph {et~al.}(2017)\citenamefont {Brown},
  \citenamefont {Mitra}, \citenamefont {Guardado-Sanchez}, \citenamefont
  {Schau{\ss}}, \citenamefont {Kondov}, \citenamefont {Khatami}, \citenamefont
  {Paiva}, \citenamefont {Trivedi}, \citenamefont {Huse},\ and\ \citenamefont
  {Bakr}}]{Brown1385}%
  \BibitemOpen
  \bibfield  {author} {\bibinfo {author} {\bibfnamefont {P.~T.}\ \bibnamefont
  {Brown}}, \bibinfo {author} {\bibfnamefont {D.}~\bibnamefont {Mitra}},
  \bibinfo {author} {\bibfnamefont {E.}~\bibnamefont {Guardado-Sanchez}},
  \bibinfo {author} {\bibfnamefont {P.}~\bibnamefont {Schau{\ss}}}, \bibinfo
  {author} {\bibfnamefont {S.~S.}\ \bibnamefont {Kondov}}, \bibinfo {author}
  {\bibfnamefont {E.}~\bibnamefont {Khatami}}, \bibinfo {author} {\bibfnamefont
  {T.}~\bibnamefont {Paiva}}, \bibinfo {author} {\bibfnamefont
  {N.}~\bibnamefont {Trivedi}}, \bibinfo {author} {\bibfnamefont {D.~A.}\
  \bibnamefont {Huse}}, \ and\ \bibinfo {author} {\bibfnamefont {W.~S.}\
  \bibnamefont {Bakr}},\ }\href {\doibase 10.1126/science.aam7838} {\bibfield
  {journal} {\bibinfo  {journal} {Science}\ }\textbf {\bibinfo {volume}
  {357}},\ \bibinfo {pages} {1385} (\bibinfo {year} {2017})},\ \Eprint
  {http://arxiv.org/abs/http://science.sciencemag.org/content/357/6358/1385.full.pdf}
  {http://science.sciencemag.org/content/357/6358/1385.full.pdf} \BibitemShut
  {NoStop}%
\bibitem [{\citenamefont {Fratino}\ \emph {et~al.}(2017)\citenamefont
  {Fratino}, \citenamefont {S\'emon}, \citenamefont {Charlebois}, \citenamefont
  {Sordi},\ and\ \citenamefont {Tremblay}}]{PhysRevB.95.235109}%
  \BibitemOpen
  \bibfield  {author} {\bibinfo {author} {\bibfnamefont {L.}~\bibnamefont
  {Fratino}}, \bibinfo {author} {\bibfnamefont {P.}~\bibnamefont {S\'emon}},
  \bibinfo {author} {\bibfnamefont {M.}~\bibnamefont {Charlebois}}, \bibinfo
  {author} {\bibfnamefont {G.}~\bibnamefont {Sordi}}, \ and\ \bibinfo {author}
  {\bibfnamefont {A.-M.~S.}\ \bibnamefont {Tremblay}},\ }\href {\doibase
  10.1103/PhysRevB.95.235109} {\bibfield  {journal} {\bibinfo  {journal} {Phys.
  Rev. B}\ }\textbf {\bibinfo {volume} {95}},\ \bibinfo {pages} {235109}
  (\bibinfo {year} {2017})}\BibitemShut {NoStop}%
\bibitem [{\citenamefont {Kumar}\ \emph {et~al.}(2016)\citenamefont {Kumar},
  \citenamefont {Mertz},\ and\ \citenamefont
  {Hofstetter}}]{PhysRevB.94.115161}%
  \BibitemOpen
  \bibfield  {author} {\bibinfo {author} {\bibfnamefont {P.}~\bibnamefont
  {Kumar}}, \bibinfo {author} {\bibfnamefont {T.}~\bibnamefont {Mertz}}, \ and\
  \bibinfo {author} {\bibfnamefont {W.}~\bibnamefont {Hofstetter}},\ }\href
  {\doibase 10.1103/PhysRevB.94.115161} {\bibfield  {journal} {\bibinfo
  {journal} {Phys. Rev. B}\ }\textbf {\bibinfo {volume} {94}},\ \bibinfo
  {pages} {115161} (\bibinfo {year} {2016})}\BibitemShut {NoStop}%
\bibitem [{\citenamefont {Heikkinen}\ \emph {et~al.}(2013)\citenamefont
  {Heikkinen}, \citenamefont {Kim},\ and\ \citenamefont
  {T\"orm\"a}}]{PhysRevB.87.224513}%
  \BibitemOpen
  \bibfield  {author} {\bibinfo {author} {\bibfnamefont {M.~O.~J.}\
  \bibnamefont {Heikkinen}}, \bibinfo {author} {\bibfnamefont {D.-H.}\
  \bibnamefont {Kim}}, \ and\ \bibinfo {author} {\bibfnamefont
  {P.}~\bibnamefont {T\"orm\"a}},\ }\href {\doibase 10.1103/PhysRevB.87.224513}
  {\bibfield  {journal} {\bibinfo  {journal} {Phys. Rev. B}\ }\textbf {\bibinfo
  {volume} {87}},\ \bibinfo {pages} {224513} (\bibinfo {year}
  {2013})}\BibitemShut {NoStop}%
\bibitem [{\citenamefont {Snoek}\ \emph {et~al.}(2011)\citenamefont {Snoek},
  \citenamefont {Titvinidze},\ and\ \citenamefont
  {Hofstetter}}]{PhysRevB.83.054419}%
  \BibitemOpen
  \bibfield  {author} {\bibinfo {author} {\bibfnamefont {M.}~\bibnamefont
  {Snoek}}, \bibinfo {author} {\bibfnamefont {I.}~\bibnamefont {Titvinidze}}, \
  and\ \bibinfo {author} {\bibfnamefont {W.}~\bibnamefont {Hofstetter}},\
  }\href {\doibase 10.1103/PhysRevB.83.054419} {\bibfield  {journal} {\bibinfo
  {journal} {Phys. Rev. B}\ }\textbf {\bibinfo {volume} {83}},\ \bibinfo
  {pages} {054419} (\bibinfo {year} {2011})}\BibitemShut {NoStop}%
\bibitem [{\citenamefont {Gull}\ \emph {et~al.}(2011)\citenamefont {Gull},
  \citenamefont {Millis}, \citenamefont {Lichtenstein}, \citenamefont
  {Rubtsov}, \citenamefont {Troyer},\ and\ \citenamefont
  {Werner}}]{RevModPhys.83.349}%
  \BibitemOpen
  \bibfield  {author} {\bibinfo {author} {\bibfnamefont {E.}~\bibnamefont
  {Gull}}, \bibinfo {author} {\bibfnamefont {A.~J.}\ \bibnamefont {Millis}},
  \bibinfo {author} {\bibfnamefont {A.~I.}\ \bibnamefont {Lichtenstein}},
  \bibinfo {author} {\bibfnamefont {A.~N.}\ \bibnamefont {Rubtsov}}, \bibinfo
  {author} {\bibfnamefont {M.}~\bibnamefont {Troyer}}, \ and\ \bibinfo {author}
  {\bibfnamefont {P.}~\bibnamefont {Werner}},\ }\href {\doibase
  10.1103/RevModPhys.83.349} {\bibfield  {journal} {\bibinfo  {journal} {Rev.
  Mod. Phys.}\ }\textbf {\bibinfo {volume} {83}},\ \bibinfo {pages} {349}
  (\bibinfo {year} {2011})}\BibitemShut {NoStop}%
\bibitem [{\citenamefont {Peters}\ and\ \citenamefont
  {Kawakami}(2014)}]{PhysRevB.89.155134}%
  \BibitemOpen
  \bibfield  {author} {\bibinfo {author} {\bibfnamefont {R.}~\bibnamefont
  {Peters}}\ and\ \bibinfo {author} {\bibfnamefont {N.}~\bibnamefont
  {Kawakami}},\ }\href {\doibase 10.1103/PhysRevB.89.155134} {\bibfield
  {journal} {\bibinfo  {journal} {Phys. Rev. B}\ }\textbf {\bibinfo {volume}
  {89}},\ \bibinfo {pages} {155134} (\bibinfo {year} {2014})}\BibitemShut
  {NoStop}%
\bibitem [{\citenamefont {Vanhala}\ and\ \citenamefont
  {T{\"o}rm{\"a}}(2017)}]{vanhala2017dynamical}%
  \BibitemOpen
  \bibfield  {author} {\bibinfo {author} {\bibfnamefont {T.~I.}\ \bibnamefont
  {Vanhala}}\ and\ \bibinfo {author} {\bibfnamefont {P.}~\bibnamefont
  {T{\"o}rm{\"a}}},\ }\href@noop {} {\bibfield  {journal} {\bibinfo  {journal}
  {arXiv preprint arXiv:1708.06749}\ } (\bibinfo {year} {2017})}\BibitemShut
  {NoStop}%
\bibitem [{\citenamefont {Georges}\ \emph {et~al.}(1996)\citenamefont
  {Georges}, \citenamefont {Kotliar}, \citenamefont {Krauth},\ and\
  \citenamefont {Rozenberg}}]{RevModPhys.68.13}%
  \BibitemOpen
  \bibfield  {author} {\bibinfo {author} {\bibfnamefont {A.}~\bibnamefont
  {Georges}}, \bibinfo {author} {\bibfnamefont {G.}~\bibnamefont {Kotliar}},
  \bibinfo {author} {\bibfnamefont {W.}~\bibnamefont {Krauth}}, \ and\ \bibinfo
  {author} {\bibfnamefont {M.~J.}\ \bibnamefont {Rozenberg}},\ }\href {\doibase
  10.1103/RevModPhys.68.13} {\bibfield  {journal} {\bibinfo  {journal} {Rev.
  Mod. Phys.}\ }\textbf {\bibinfo {volume} {68}},\ \bibinfo {pages} {13}
  (\bibinfo {year} {1996})}\BibitemShut {NoStop}%
\bibitem [{\citenamefont {Snoek}\ \emph {et~al.}(2008)\citenamefont {Snoek},
  \citenamefont {Titvinidze}, \citenamefont {T\"oke}, \citenamefont {Byczuk},\
  and\ \citenamefont {Hofstetter}}]{1367-2630-10-9-093008}%
  \BibitemOpen
  \bibfield  {author} {\bibinfo {author} {\bibfnamefont {M.}~\bibnamefont
  {Snoek}}, \bibinfo {author} {\bibfnamefont {I.}~\bibnamefont {Titvinidze}},
  \bibinfo {author} {\bibfnamefont {C.}~\bibnamefont {T\"oke}}, \bibinfo
  {author} {\bibfnamefont {K.}~\bibnamefont {Byczuk}}, \ and\ \bibinfo {author}
  {\bibfnamefont {W.}~\bibnamefont {Hofstetter}},\ }\href
  {http://stacks.iop.org/1367-2630/10/i=9/a=093008} {\bibfield  {journal}
  {\bibinfo  {journal} {New Journal of Physics}\ }\textbf {\bibinfo {volume}
  {10}},\ \bibinfo {pages} {093008} (\bibinfo {year} {2008})}\BibitemShut
  {NoStop}%
\bibitem [{\citenamefont {Maier}\ \emph {et~al.}(2005)\citenamefont {Maier},
  \citenamefont {Jarrell}, \citenamefont {Pruschke},\ and\ \citenamefont
  {Hettler}}]{RevModPhys.77.1027}%
  \BibitemOpen
  \bibfield  {author} {\bibinfo {author} {\bibfnamefont {T.}~\bibnamefont
  {Maier}}, \bibinfo {author} {\bibfnamefont {M.}~\bibnamefont {Jarrell}},
  \bibinfo {author} {\bibfnamefont {T.}~\bibnamefont {Pruschke}}, \ and\
  \bibinfo {author} {\bibfnamefont {M.~H.}\ \bibnamefont {Hettler}},\ }\href
  {\doibase 10.1103/RevModPhys.77.1027} {\bibfield  {journal} {\bibinfo
  {journal} {Rev. Mod. Phys.}\ }\textbf {\bibinfo {volume} {77}},\ \bibinfo
  {pages} {1027} (\bibinfo {year} {2005})}\BibitemShut {NoStop}%
\bibitem [{\citenamefont {Jordens}\ \emph {et~al.}(2008)\citenamefont
  {Jordens}, \citenamefont {Strohmaier}, \citenamefont {Gunter}, \citenamefont
  {Moritz},\ and\ \citenamefont {Esslinger}}]{Jordens2008}%
  \BibitemOpen
  \bibfield  {author} {\bibinfo {author} {\bibfnamefont {R.}~\bibnamefont
  {Jordens}}, \bibinfo {author} {\bibfnamefont {N.}~\bibnamefont {Strohmaier}},
  \bibinfo {author} {\bibfnamefont {K.}~\bibnamefont {Gunter}}, \bibinfo
  {author} {\bibfnamefont {H.}~\bibnamefont {Moritz}}, \ and\ \bibinfo {author}
  {\bibfnamefont {T.}~\bibnamefont {Esslinger}},\ }\href {\doibase
  10.1038/nature07244} {\bibfield  {journal} {\bibinfo  {journal} {Nature}\
  }\textbf {\bibinfo {volume} {455}},\ \bibinfo {pages} {204} (\bibinfo {year}
  {2008})}\BibitemShut {NoStop}%
\bibitem [{\citenamefont {Kune\ifmmode~\check{s}\else
  \v{s}\fi{}}(2011)}]{PhysRevB.83.085102}%
  \BibitemOpen
  \bibfield  {author} {\bibinfo {author} {\bibfnamefont {J.}~\bibnamefont
  {Kune\ifmmode~\check{s}\else \v{s}\fi{}}},\ }\href {\doibase
  10.1103/PhysRevB.83.085102} {\bibfield  {journal} {\bibinfo  {journal} {Phys.
  Rev. B}\ }\textbf {\bibinfo {volume} {83}},\ \bibinfo {pages} {085102}
  (\bibinfo {year} {2011})}\BibitemShut {NoStop}%
\bibitem [{\citenamefont {Kopnin}\ \emph {et~al.}(2011)\citenamefont {Kopnin},
  \citenamefont {Heikkil\"a},\ and\ \citenamefont
  {Volovik}}]{PhysRevB.83.220503}%
  \BibitemOpen
  \bibfield  {author} {\bibinfo {author} {\bibfnamefont {N.~B.}\ \bibnamefont
  {Kopnin}}, \bibinfo {author} {\bibfnamefont {T.~T.}\ \bibnamefont
  {Heikkil\"a}}, \ and\ \bibinfo {author} {\bibfnamefont {G.~E.}\ \bibnamefont
  {Volovik}},\ }\href {\doibase 10.1103/PhysRevB.83.220503} {\bibfield
  {journal} {\bibinfo  {journal} {Phys. Rev. B}\ }\textbf {\bibinfo {volume}
  {83}},\ \bibinfo {pages} {220503} (\bibinfo {year} {2011})}\BibitemShut
  {NoStop}%
\bibitem [{\citenamefont {Heikkil{\"a}}\ \emph {et~al.}(2011)\citenamefont
  {Heikkil{\"a}}, \citenamefont {Kopnin},\ and\ \citenamefont
  {Volovik}}]{Heikkilä2011}%
  \BibitemOpen
  \bibfield  {author} {\bibinfo {author} {\bibfnamefont {T.~T.}\ \bibnamefont
  {Heikkil{\"a}}}, \bibinfo {author} {\bibfnamefont {N.~B.}\ \bibnamefont
  {Kopnin}}, \ and\ \bibinfo {author} {\bibfnamefont {G.~E.}\ \bibnamefont
  {Volovik}},\ }\href {\doibase 10.1134/S0021364011150045} {\bibfield
  {journal} {\bibinfo  {journal} {JETP Letters}\ }\textbf {\bibinfo {volume}
  {94}},\ \bibinfo {pages} {233} (\bibinfo {year} {2011})}\BibitemShut
  {NoStop}%
\bibitem [{\citenamefont {Hohenadler}\ \emph {et~al.}(2014)\citenamefont
  {Hohenadler}, \citenamefont {Parisen~Toldin}, \citenamefont {Herbut},\ and\
  \citenamefont {Assaad}}]{PhysRevB.90.085146}%
  \BibitemOpen
  \bibfield  {author} {\bibinfo {author} {\bibfnamefont {M.}~\bibnamefont
  {Hohenadler}}, \bibinfo {author} {\bibfnamefont {F.}~\bibnamefont
  {Parisen~Toldin}}, \bibinfo {author} {\bibfnamefont {I.~F.}\ \bibnamefont
  {Herbut}}, \ and\ \bibinfo {author} {\bibfnamefont {F.~F.}\ \bibnamefont
  {Assaad}},\ }\href {\doibase 10.1103/PhysRevB.90.085146} {\bibfield
  {journal} {\bibinfo  {journal} {Phys. Rev. B}\ }\textbf {\bibinfo {volume}
  {90}},\ \bibinfo {pages} {085146} (\bibinfo {year} {2014})}\BibitemShut
  {NoStop}%
\bibitem [{\citenamefont {Gull}\ \emph {et~al.}(2008)\citenamefont {Gull},
  \citenamefont {Werner}, \citenamefont {Wang}, \citenamefont {Troyer},\ and\
  \citenamefont {Millis}}]{0295-5075-84-3-37009}%
  \BibitemOpen
  \bibfield  {author} {\bibinfo {author} {\bibfnamefont {E.}~\bibnamefont
  {Gull}}, \bibinfo {author} {\bibfnamefont {P.}~\bibnamefont {Werner}},
  \bibinfo {author} {\bibfnamefont {X.}~\bibnamefont {Wang}}, \bibinfo {author}
  {\bibfnamefont {M.}~\bibnamefont {Troyer}}, \ and\ \bibinfo {author}
  {\bibfnamefont {A.~J.}\ \bibnamefont {Millis}},\ }\href
  {http://stacks.iop.org/0295-5075/84/i=3/a=37009} {\bibfield  {journal}
  {\bibinfo  {journal} {EPL (Europhysics Letters)}\ }\textbf {\bibinfo {volume}
  {84}},\ \bibinfo {pages} {37009} (\bibinfo {year} {2008})}\BibitemShut
  {NoStop}%
\bibitem [{\citenamefont {Gopalan}\ \emph {et~al.}(1992)\citenamefont
  {Gopalan}, \citenamefont {Gunnarsson},\ and\ \citenamefont
  {Andersen}}]{PhysRevB.46.11798}%
  \BibitemOpen
  \bibfield  {author} {\bibinfo {author} {\bibfnamefont {S.}~\bibnamefont
  {Gopalan}}, \bibinfo {author} {\bibfnamefont {O.}~\bibnamefont {Gunnarsson}},
  \ and\ \bibinfo {author} {\bibfnamefont {O.~K.}\ \bibnamefont {Andersen}},\
  }\href {\doibase 10.1103/PhysRevB.46.11798} {\bibfield  {journal} {\bibinfo
  {journal} {Phys. Rev. B}\ }\textbf {\bibinfo {volume} {46}},\ \bibinfo
  {pages} {11798} (\bibinfo {year} {1992})}\BibitemShut {NoStop}%
\bibitem [{\citenamefont {Hlubina}\ and\ \citenamefont
  {Rice}(1995)}]{PhysRevB.51.9253}%
  \BibitemOpen
  \bibfield  {author} {\bibinfo {author} {\bibfnamefont {R.}~\bibnamefont
  {Hlubina}}\ and\ \bibinfo {author} {\bibfnamefont {T.~M.}\ \bibnamefont
  {Rice}},\ }\href {\doibase 10.1103/PhysRevB.51.9253} {\bibfield  {journal}
  {\bibinfo  {journal} {Phys. Rev. B}\ }\textbf {\bibinfo {volume} {51}},\
  \bibinfo {pages} {9253} (\bibinfo {year} {1995})}\BibitemShut {NoStop}%
\bibitem [{\citenamefont {{Igor Dzyaloshinskii}}(1996)}]{refId0}%
  \BibitemOpen
  \bibfield  {author} {\bibinfo {author} {\bibnamefont {{Igor
  Dzyaloshinskii}}},\ }\href {\doibase 10.1051/jp1:1996127} {\bibfield
  {journal} {\bibinfo  {journal} {J. Phys. I France}\ }\textbf {\bibinfo
  {volume} {6}},\ \bibinfo {pages} {119} (\bibinfo {year} {1996})}\BibitemShut
  {NoStop}%
\bibitem [{\citenamefont {Katanin}\ and\ \citenamefont
  {Kampf}(2003)}]{PhysRevB.68.195101}%
  \BibitemOpen
  \bibfield  {author} {\bibinfo {author} {\bibfnamefont {A.~A.}\ \bibnamefont
  {Katanin}}\ and\ \bibinfo {author} {\bibfnamefont {A.~P.}\ \bibnamefont
  {Kampf}},\ }\href {\doibase 10.1103/PhysRevB.68.195101} {\bibfield  {journal}
  {\bibinfo  {journal} {Phys. Rev. B}\ }\textbf {\bibinfo {volume} {68}},\
  \bibinfo {pages} {195101} (\bibinfo {year} {2003})}\BibitemShut {NoStop}%
\bibitem [{\citenamefont {Varma}\ \emph {et~al.}(1989)\citenamefont {Varma},
  \citenamefont {Littlewood}, \citenamefont {Schmitt-Rink}, \citenamefont
  {Abrahams},\ and\ \citenamefont {Ruckenstein}}]{PhysRevLett.63.1996}%
  \BibitemOpen
  \bibfield  {author} {\bibinfo {author} {\bibfnamefont {C.~M.}\ \bibnamefont
  {Varma}}, \bibinfo {author} {\bibfnamefont {P.~B.}\ \bibnamefont
  {Littlewood}}, \bibinfo {author} {\bibfnamefont {S.}~\bibnamefont
  {Schmitt-Rink}}, \bibinfo {author} {\bibfnamefont {E.}~\bibnamefont
  {Abrahams}}, \ and\ \bibinfo {author} {\bibfnamefont {A.~E.}\ \bibnamefont
  {Ruckenstein}},\ }\href {\doibase 10.1103/PhysRevLett.63.1996} {\bibfield
  {journal} {\bibinfo  {journal} {Phys. Rev. Lett.}\ }\textbf {\bibinfo
  {volume} {63}},\ \bibinfo {pages} {1996} (\bibinfo {year}
  {1989})}\BibitemShut {NoStop}%
\bibitem [{\citenamefont {Schmitt}(2010)}]{PhysRevB.82.155126}%
  \BibitemOpen
  \bibfield  {author} {\bibinfo {author} {\bibfnamefont {S.}~\bibnamefont
  {Schmitt}},\ }\href {\doibase 10.1103/PhysRevB.82.155126} {\bibfield
  {journal} {\bibinfo  {journal} {Phys. Rev. B}\ }\textbf {\bibinfo {volume}
  {82}},\ \bibinfo {pages} {155126} (\bibinfo {year} {2010})}\BibitemShut
  {NoStop}%
\bibitem [{\citenamefont {Rubtsov}\ \emph {et~al.}(2009)\citenamefont
  {Rubtsov}, \citenamefont {Katsnelson}, \citenamefont {Lichtenstein},\ and\
  \citenamefont {Georges}}]{PhysRevB.79.045133}%
  \BibitemOpen
  \bibfield  {author} {\bibinfo {author} {\bibfnamefont {A.~N.}\ \bibnamefont
  {Rubtsov}}, \bibinfo {author} {\bibfnamefont {M.~I.}\ \bibnamefont
  {Katsnelson}}, \bibinfo {author} {\bibfnamefont {A.~I.}\ \bibnamefont
  {Lichtenstein}}, \ and\ \bibinfo {author} {\bibfnamefont {A.}~\bibnamefont
  {Georges}},\ }\href {\doibase 10.1103/PhysRevB.79.045133} {\bibfield
  {journal} {\bibinfo  {journal} {Phys. Rev. B}\ }\textbf {\bibinfo {volume}
  {79}},\ \bibinfo {pages} {045133} (\bibinfo {year} {2009})}\BibitemShut
  {NoStop}%
\bibitem [{\citenamefont {Liebsch}\ and\ \citenamefont
  {Tong}(2009)}]{PhysRevB.80.165126}%
  \BibitemOpen
  \bibfield  {author} {\bibinfo {author} {\bibfnamefont {A.}~\bibnamefont
  {Liebsch}}\ and\ \bibinfo {author} {\bibfnamefont {N.-H.}\ \bibnamefont
  {Tong}},\ }\href {\doibase 10.1103/PhysRevB.80.165126} {\bibfield  {journal}
  {\bibinfo  {journal} {Phys. Rev. B}\ }\textbf {\bibinfo {volume} {80}},\
  \bibinfo {pages} {165126} (\bibinfo {year} {2009})}\BibitemShut {NoStop}%
\bibitem [{\citenamefont {Amaricci}\ \emph {et~al.}(2008)\citenamefont
  {Amaricci}, \citenamefont {Sordi},\ and\ \citenamefont
  {Rozenberg}}]{PhysRevLett.101.146403}%
  \BibitemOpen
  \bibfield  {author} {\bibinfo {author} {\bibfnamefont {A.}~\bibnamefont
  {Amaricci}}, \bibinfo {author} {\bibfnamefont {G.}~\bibnamefont {Sordi}}, \
  and\ \bibinfo {author} {\bibfnamefont {M.~J.}\ \bibnamefont {Rozenberg}},\
  }\href {\doibase 10.1103/PhysRevLett.101.146403} {\bibfield  {journal}
  {\bibinfo  {journal} {Phys. Rev. Lett.}\ }\textbf {\bibinfo {volume} {101}},\
  \bibinfo {pages} {146403} (\bibinfo {year} {2008})}\BibitemShut {NoStop}%
\bibitem [{\citenamefont {Tovmasyan}\ \emph {et~al.}(2016)\citenamefont
  {Tovmasyan}, \citenamefont {Peotta}, \citenamefont {T\"orm\"a},\ and\
  \citenamefont {Huber}}]{PhysRevB.94.245149}%
  \BibitemOpen
  \bibfield  {author} {\bibinfo {author} {\bibfnamefont {M.}~\bibnamefont
  {Tovmasyan}}, \bibinfo {author} {\bibfnamefont {S.}~\bibnamefont {Peotta}},
  \bibinfo {author} {\bibfnamefont {P.}~\bibnamefont {T\"orm\"a}}, \ and\
  \bibinfo {author} {\bibfnamefont {S.~D.}\ \bibnamefont {Huber}},\ }\href
  {\doibase 10.1103/PhysRevB.94.245149} {\bibfield  {journal} {\bibinfo
  {journal} {Phys. Rev. B}\ }\textbf {\bibinfo {volume} {94}},\ \bibinfo
  {pages} {245149} (\bibinfo {year} {2016})}\BibitemShut {NoStop}%
\bibitem [{\citenamefont {Assaad}\ and\ \citenamefont
  {Lang}(2007)}]{PhysRevB.76.035116}%
  \BibitemOpen
  \bibfield  {author} {\bibinfo {author} {\bibfnamefont {F.~F.}\ \bibnamefont
  {Assaad}}\ and\ \bibinfo {author} {\bibfnamefont {T.~C.}\ \bibnamefont
  {Lang}},\ }\href {\doibase 10.1103/PhysRevB.76.035116} {\bibfield  {journal}
  {\bibinfo  {journal} {Phys. Rev. B}\ }\textbf {\bibinfo {volume} {76}},\
  \bibinfo {pages} {035116} (\bibinfo {year} {2007})}\BibitemShut {NoStop}%
\bibitem [{\citenamefont {Costa}\ \emph {et~al.}(2016)\citenamefont {Costa},
  \citenamefont {Mendes-Santos}, \citenamefont {Paiva}, \citenamefont
  {Santos},\ and\ \citenamefont {Scalettar}}]{PhysRevB.94.155107}%
  \BibitemOpen
  \bibfield  {author} {\bibinfo {author} {\bibfnamefont {N.~C.}\ \bibnamefont
  {Costa}}, \bibinfo {author} {\bibfnamefont {T.}~\bibnamefont
  {Mendes-Santos}}, \bibinfo {author} {\bibfnamefont {T.}~\bibnamefont
  {Paiva}}, \bibinfo {author} {\bibfnamefont {R.~R.~d.}\ \bibnamefont
  {Santos}}, \ and\ \bibinfo {author} {\bibfnamefont {R.~T.}\ \bibnamefont
  {Scalettar}},\ }\href {\doibase 10.1103/PhysRevB.94.155107} {\bibfield
  {journal} {\bibinfo  {journal} {Phys. Rev. B}\ }\textbf {\bibinfo {volume}
  {94}},\ \bibinfo {pages} {155107} (\bibinfo {year} {2016})}\BibitemShut
  {NoStop}%
\bibitem [{\citenamefont {Xu}\ \emph {et~al.}(2011)\citenamefont {Xu},
  \citenamefont {Chang}, \citenamefont {Walter},\ and\ \citenamefont
  {Zhang}}]{0953-8984-23-50-505601}%
  \BibitemOpen
  \bibfield  {author} {\bibinfo {author} {\bibfnamefont {J.}~\bibnamefont
  {Xu}}, \bibinfo {author} {\bibfnamefont {C.-C.}\ \bibnamefont {Chang}},
  \bibinfo {author} {\bibfnamefont {E.~J.}\ \bibnamefont {Walter}}, \ and\
  \bibinfo {author} {\bibfnamefont {S.}~\bibnamefont {Zhang}},\ }\href
  {http://stacks.iop.org/0953-8984/23/i=50/a=505601} {\bibfield  {journal}
  {\bibinfo  {journal} {Journal of Physics: Condensed Matter}\ }\textbf
  {\bibinfo {volume} {23}},\ \bibinfo {pages} {505601} (\bibinfo {year}
  {2011})}\BibitemShut {NoStop}%
\bibitem [{\citenamefont {Fink}\ \emph {et~al.}(2009)\citenamefont {Fink},
  \citenamefont {Schierle}, \citenamefont {Weschke}, \citenamefont {Geck},
  \citenamefont {Hawthorn}, \citenamefont {Soltwisch}, \citenamefont {Wadati},
  \citenamefont {Wu}, \citenamefont {D\"urr}, \citenamefont {Wizent},
  \citenamefont {B\"uchner},\ and\ \citenamefont
  {Sawatzky}}]{PhysRevB.79.100502}%
  \BibitemOpen
  \bibfield  {author} {\bibinfo {author} {\bibfnamefont {J.}~\bibnamefont
  {Fink}}, \bibinfo {author} {\bibfnamefont {E.}~\bibnamefont {Schierle}},
  \bibinfo {author} {\bibfnamefont {E.}~\bibnamefont {Weschke}}, \bibinfo
  {author} {\bibfnamefont {J.}~\bibnamefont {Geck}}, \bibinfo {author}
  {\bibfnamefont {D.}~\bibnamefont {Hawthorn}}, \bibinfo {author}
  {\bibfnamefont {V.}~\bibnamefont {Soltwisch}}, \bibinfo {author}
  {\bibfnamefont {H.}~\bibnamefont {Wadati}}, \bibinfo {author} {\bibfnamefont
  {H.-H.}\ \bibnamefont {Wu}}, \bibinfo {author} {\bibfnamefont {H.~A.}\
  \bibnamefont {D\"urr}}, \bibinfo {author} {\bibfnamefont {N.}~\bibnamefont
  {Wizent}}, \bibinfo {author} {\bibfnamefont {B.}~\bibnamefont {B\"uchner}}, \
  and\ \bibinfo {author} {\bibfnamefont {G.~A.}\ \bibnamefont {Sawatzky}},\
  }\href {\doibase 10.1103/PhysRevB.79.100502} {\bibfield  {journal} {\bibinfo
  {journal} {Phys. Rev. B}\ }\textbf {\bibinfo {volume} {79}},\ \bibinfo
  {pages} {100502} (\bibinfo {year} {2009})}\BibitemShut {NoStop}%
\bibitem [{\citenamefont {Kinnunen}\ \emph {et~al.}(2017)\citenamefont
  {Kinnunen}, \citenamefont {Baarsma}, \citenamefont {Martikainen},\ and\
  \citenamefont {T{\"o}rm{\"a}}}]{kinnunen2017fulde}%
  \BibitemOpen
  \bibfield  {author} {\bibinfo {author} {\bibfnamefont {J.~J.}\ \bibnamefont
  {Kinnunen}}, \bibinfo {author} {\bibfnamefont {J.~E.}\ \bibnamefont
  {Baarsma}}, \bibinfo {author} {\bibfnamefont {J.-P.}\ \bibnamefont
  {Martikainen}}, \ and\ \bibinfo {author} {\bibfnamefont {P.}~\bibnamefont
  {T{\"o}rm{\"a}}},\ }\href@noop {} {\bibfield  {journal} {\bibinfo  {journal}
  {arXiv preprint arXiv:1706.07076}\ } (\bibinfo {year} {2017})}\BibitemShut
  {NoStop}%
\bibitem [{\citenamefont {Tranquada}(2013)}]{doi:10.1063/1.4818402}%
  \BibitemOpen
  \bibfield  {author} {\bibinfo {author} {\bibfnamefont {J.~M.}\ \bibnamefont
  {Tranquada}},\ }\href {\doibase 10.1063/1.4818402} {\bibfield  {journal}
  {\bibinfo  {journal} {AIP Conference Proceedings}\ }\textbf {\bibinfo
  {volume} {1550}},\ \bibinfo {pages} {114} (\bibinfo {year} {2013})},\ \Eprint
  {http://arxiv.org/abs/http://aip.scitation.org/doi/pdf/10.1063/1.4818402}
  {http://aip.scitation.org/doi/pdf/10.1063/1.4818402} \BibitemShut {NoStop}%
\bibitem [{\citenamefont {Corboz}\ \emph {et~al.}(2014)\citenamefont {Corboz},
  \citenamefont {Rice},\ and\ \citenamefont {Troyer}}]{PhysRevLett.113.046402}%
  \BibitemOpen
  \bibfield  {author} {\bibinfo {author} {\bibfnamefont {P.}~\bibnamefont
  {Corboz}}, \bibinfo {author} {\bibfnamefont {T.~M.}\ \bibnamefont {Rice}}, \
  and\ \bibinfo {author} {\bibfnamefont {M.}~\bibnamefont {Troyer}},\ }\href
  {\doibase 10.1103/PhysRevLett.113.046402} {\bibfield  {journal} {\bibinfo
  {journal} {Phys. Rev. Lett.}\ }\textbf {\bibinfo {volume} {113}},\ \bibinfo
  {pages} {046402} (\bibinfo {year} {2014})}\BibitemShut {NoStop}%
\bibitem [{\citenamefont {Raczkowski}\ and\ \citenamefont
  {Assaad}(2010)}]{PhysRevB.82.233101}%
  \BibitemOpen
  \bibfield  {author} {\bibinfo {author} {\bibfnamefont {M.}~\bibnamefont
  {Raczkowski}}\ and\ \bibinfo {author} {\bibfnamefont {F.~F.}\ \bibnamefont
  {Assaad}},\ }\href {\doibase 10.1103/PhysRevB.82.233101} {\bibfield
  {journal} {\bibinfo  {journal} {Phys. Rev. B}\ }\textbf {\bibinfo {volume}
  {82}},\ \bibinfo {pages} {233101} (\bibinfo {year} {2010})}\BibitemShut
  {NoStop}%
\bibitem [{\citenamefont {Ho}\ \emph {et~al.}(2009)\citenamefont {Ho},
  \citenamefont {Cazalilla},\ and\ \citenamefont
  {Giamarchi}}]{PhysRevA.79.033620}%
  \BibitemOpen
  \bibfield  {author} {\bibinfo {author} {\bibfnamefont {A.~F.}\ \bibnamefont
  {Ho}}, \bibinfo {author} {\bibfnamefont {M.~A.}\ \bibnamefont {Cazalilla}}, \
  and\ \bibinfo {author} {\bibfnamefont {T.}~\bibnamefont {Giamarchi}},\ }\href
  {\doibase 10.1103/PhysRevA.79.033620} {\bibfield  {journal} {\bibinfo
  {journal} {Phys. Rev. A}\ }\textbf {\bibinfo {volume} {79}},\ \bibinfo
  {pages} {033620} (\bibinfo {year} {2009})}\BibitemShut {NoStop}%
\bibitem [{\citenamefont {Moreo}\ and\ \citenamefont
  {Scalapino}(2007)}]{PhysRevLett.98.216402}%
  \BibitemOpen
  \bibfield  {author} {\bibinfo {author} {\bibfnamefont {A.}~\bibnamefont
  {Moreo}}\ and\ \bibinfo {author} {\bibfnamefont {D.~J.}\ \bibnamefont
  {Scalapino}},\ }\href {\doibase 10.1103/PhysRevLett.98.216402} {\bibfield
  {journal} {\bibinfo  {journal} {Phys. Rev. Lett.}\ }\textbf {\bibinfo
  {volume} {98}},\ \bibinfo {pages} {216402} (\bibinfo {year}
  {2007})}\BibitemShut {NoStop}%
\end{thebibliography}%
\end{document}